\documentclass{jfm}

\usepackage{amsmath}    
\usepackage{amssymb}
\usepackage{graphicx}   
\usepackage{verbatim}   
\usepackage{color}      
\usepackage{subfigure}  
\usepackage{hyperref}   
\usepackage{tensor}
\usepackage{amsfonts}
\usepackage{natbib}

\newcommand{\bnabla}{\boldsymbol{\nabla}}
\newcommand{\gsim}{\lower.7ex\hbox{$\;\stackrel{\textstyle>}{\sim}\;$}}
\newcommand{\lsim}{\lower.7ex\hbox{$\;\stackrel{\textstyle<}{\sim}\;$}}
\newcommand{\la}{\left<}
\newcommand{\ra}{\right>}
\newcommand{\bphi}{{\boldsymbol \phi}}
\newcommand{\bxi}{{\boldsymbol \xi}}
\newcommand{\Kabs}{{\cal K}_\text{abs}}

\title[Two-dimensionalization of low-$Ro$ turbulence ]{Exact two-dimensionalization of rapidly rotating large-Reynolds-number flows}

\author{Basile Gallet}
\affiliation{Service de Physique de l'\'Etat Condens\'e, CEA, CNRS, Universit\'e Paris-Saclay, CEA Saclay, 91191 Gif-sur-Yvette, France}

\begin{document}

\maketitle

\begin{abstract}
We consider the flow of a Newtonian fluid in a three-dimensional domain, rotating about a vertical axis and driven by a vertically invariant horizontal body-force. This system admits vertically invariant solutions that satisfy the 2D Navier-Stokes equation. At high Reynolds number and without global rotation, such solutions are usually unstable to three-dimensional perturbations. By contrast, for strong enough global rotation, we prove rigorously that the 2D (and possibly turbulent) solutions are stable to vertically dependent perturbations. 

We first consider the 3D rotating Navier-Stokes equation linearized around a statistically steady 2D flow solution. We show that this base flow is linearly stable to vertically dependent perturbations when the global rotation is fast enough: under a Reynolds-number-dependent threshold value $Ro_c(Re)$ of the Rossby number, the flow becomes exactly 2D in the long-time limit, provided that the initial 3D perturbations are small. We call this property {\it linear two-dimensionalization}. We compute explicit lower bounds on $Ro_c(Re)$ and therefore determine regions of the parameter space $(Re,Ro)$ where such exact two-dimensionalization takes place. We present similar results in terms of the forcing strength instead of the root-mean-square velocity: the global attractor of the 2D Navier-Stokes equation is linearly stable to vertically dependent perturbations when the forcing-based Rossby number $Ro^{(f)}$ is lower than a Grashof-number-dependent threshold value $Ro^{(f)}_c(Gr)$.

We then consider the fully nonlinear 3D rotating Navier-Stokes equation and prove {\it absolute two-dimensionalization}: 
we show that, below some threshold value $Ro^{(f)}_{\text{abs}}(Gr)$ of the forcing-based Rossby number, the flow becomes two-dimensional in the long-time limit, regardless of the initial condition (including initial 3D perturbations of arbitrarily large amplitude). 

These results shed some light on several fundamental questions of rotating turbulence: for arbitrary Reynolds number $Re$ and small enough Rossby number, the system is attracted towards purely 2D flow solutions, which display no energy dissipation anomaly and no cyclone-anticyclone asymmetry. Finally, these results challenge the applicability of wave turbulence theory to describe stationary rotating turbulence in bounded domains.
\end{abstract}

\section{Introduction}

Global rotation is ubiquitous in geophysical, astrophysical and industrial flows. Uniform solid body rotation at angular frequency $\Omega$ affects the fluid motion through the action of the Coriolis force, and allows for inertial waves: in an inviscid and incompressible fluid, an infinitesimal wave-like velocity disturbance obeys the inertial-wave dispersion relation,
\begin{equation}
\sigma=\pm2 \Omega \frac{k_z}{k} \, ,
\end{equation}
where $\sigma$ is the angular frequency, ${\bf k}$ is the wave vector, $k=|{\bf k}|$, and $k_z$ is the component of the wave vector along the axis of global rotation (denoted as the vertical $z$-axis by convention).

Both the linear and fully nonlinear behaviors of the flow are therefore affected by global rotation. For turbulent flows, the strength of global rotation can be characterized by the Rossby number $Ro$, defined as the ratio of the global rotation period to the large-scale eddy turnover time. 
When the Rossby number is low, global rotation induces strong anisotropy: the flow tends to become two-dimensional, with flow structures weakly dependent on the coordinate along the rotation axis \citep{DavidsonBook}. 
This result is usually referred to as Taylor-Proudman theorem, which considers the asymptotic limit of vanishing Rossby number (infinite global rotation rate): fluid motion with characteristic time much larger than the rotation period is independent of the vertical. 

Turbulent flows at large Reynolds number $Re$ contain a broad range of spatial scales and temporal frequencies, including frequencies very large compared with the inverse large-scale eddy turnover time. While the large-scale and low-frequency structures of the flow become 2D for strong enough global rotation, the fate of small-scale high-frequency structures is less clear, and whether the latter become 2D as well for rapid global rotation is an open issue of rotating turbulence. This constitutes the central question of this study: are rotating flows more and more 2D as $Ro$ decreases, with a nonzero but decreasing fraction of the total energy contained in fully 3D flow structures, or do they become {\it exactly} two-dimensional under a critical value of the Rossby number, with no dependence at all along the vertical?

This central question is related to many of the fundamental questions addressed by experimental and numerical studies on rotating turbulence:
\begin{itemize}
\item How much power per unit mass $\epsilon$ does a rotating turbulent flow dissipate? For stationary rotating turbulence with root-mean-square velocity $U$ and length scale $\ell$, does $\epsilon$ display a dissipation anomaly, with $\lim_{Re \to \infty} \frac{\epsilon \ell}{U^3} > 0$, like in classical 3D turbulence \citep{Frisch, Doering}, or does it behave like 2D flows, with $\lim_{Re \to \infty} \frac{\epsilon \ell}{U^3} = 0$ \citep{AlexakisDoering}?
\item Why does rotating turbulence display less intermittency than its non-rotating counterpart \citep{Baroud, Muller, Seiwert, Mininni}?
\item Global rotation induces an asymmetry of the vertical vorticity distribution. Such cyclone-anticyclone asymmetry is observed in experimental and numerical studies at moderately low values of the Rossby number \citep{Bartello, Bourouiba, SmithWaleffe, Morize, Sreenivasan, Moisy, Deusebio, Gallet2014, Naso}. Does cyclone-anticyclone asymmetry persist for very low Rossby number, or is it a finite-Rossby-number effect?
\item Can low-Rossby-number rotating turbulence be described in the framework of weak turbulence of inertial waves \citep{Galtier, Cambon, Yarom, Scott}?
\end{itemize}

Although these questions have been thoroughly addressed experimentally and numerically, exact mathematical results on this matter are scarce. Such exact results can be very valuable to test the various rotating turbulence models that have been proposed (see for instance \citet{CambonBook}): the model has to be compatible with the exact mathematical result in the range of parameters where the latter is valid.

In this paper we use rigorous analysis and estimates to answer the central question raised above: we consider the flow of a Newtonian fluid driven by a vertically invariant horizontal body force, and subject to steady global rotation about the vertical axis. We focus on domains that are periodic in the horizontal and bounded vertically by stress-free surfaces, although the results carry over to domains that are periodic in the three directions. Such boundary conditions are fashionable among numericists and more amenable to analysis than more realistic domains with no-slip boundary conditions. The system admits purely 2D (vertically invariant) solutions, either laminar or turbulent. In the absence of global rotation, such solutions are usually unstable to vertically dependent perturbations, so the flow is fully three-dimensional. By contrast, here we prove that, for strong enough global rotation, the 2D flow solutions are stable with respect to three-dimensional perturbations.

We first consider infinitesimal vertically dependent perturbations on a statistically-steady 2D base flow and prove {\it linear two-dimensionalization}: using a Reynolds number $Re$ and a Rossby number $Ro$ based on the r.m.s. velocity (see section \ref{2Dsol} for the exact definitions), we show that, for any given value of $Re$, there is a critical value $Ro_c(Re)$ of the Rossby number under which the -- possibly turbulent -- 2D flow is linearly stable to 3D perturbations. We compute some lower bounds on $Ro_c(Re)$ and we therefore determine a region of the parameter space $(Re,Ro)$ where such {\it exact two-dimensionalization} takes place. For generic time-independent forcing the lower bound $Ro_<$ on $Ro_c$ is given by (\ref{conditionRo}); it scales as $Ro_< \sim Re^{-6} \ln^{-2} \left( Re \right) \times (L/H)^3$, where $H/L$ is the vertical aspect ratio of the domain. This bound can be slightly improved if the forcing is of ``single-mode'' type, i.e., if it contains a single wavenumber (see equation (\ref{conditionRoKolmogorov})). We obtain similar results in terms of dimensionless numbers that involve the forcing strength instead of the r.m.s. velocity (exact definitions in section \ref{2Dsol}): when the forcing-based Rossby number $Ro^{(f)}$ is lower than a Grashof-number-dependent threshold value $Ro^{(f)}_c(Gr)$, the global attractor of the 2D Navier-Stokes equation is linearly stable to 3D perturbations. We determine regions of the parameter space $(Gr, Ro^{(f)})$ where exact two-dimensionalization takes place by deriving a lower bound $Ro^{(f)}_<$ on $Ro^{(f)}_c$, given by expression ({\ref{conditionRoF}}). It scales as $Ro^{(f)}_< \sim Gr^{-7/2} \ln^{-2} \left( Gr \right) \times (L/H)^3$.

We then consider the fully nonlinear rotating 3D Navier-Stokes equation, with arbitrarily large initial 3D velocity perturbations. Using a theorem from \citet{Babin2} on the existence of a global attractor for the 3D rotating Navier-Stokes equation, we prove {\it absolute two-dimensionalization}: when the forcing-based Rossby number is lower than a threshold value $Ro^{(f)}_{\text{abs}}(Gr)$, the flow becomes 2D in the long-time limit, regardless of the initial condition. This indicates that the global attractor of the 2D Navier-Stokes equation is the only attractor of the 3D rotating Navier-Stokes equation when $Ro^{(f)}<Ro^{(f)}_{\text{abs}}(Gr)$.

The analysis consists in studying the stability of a (possibly turbulent) 2D base flow to vertically dependent 3D perturbations. The procedure is very different from a usual stability analysis, because we do not know the exact expression for this base flow, nor do we know its precise spatial and temporal dependence. The proofs therefore rely on rigorous upper bounds for several quantities associated with such 2D flows. These bounds provide sufficient information to determine regions of the $(Re,Ro)$ parameter space where the 2D flow is stable to fully 3D perturbations.

 A somewhat similar stability analysis of a possibly turbulent 2D base flow was performed in \citet{Gallet2015}, in the context of low-magnetic-Reynolds-number ($Rm$) magnetohydrodynamic (MHD) turbulence subject to a strong external magnetic field. However, we stress the fact that the proof of stability is very different in the two situations: in the low-$Rm$ MHD case, the proof relies on Ohmic dissipation strongly damping the 3D perturbations and therefore stabilizing the 2D flow. By contrast, for rotating flows the Coriolis force does not do work and global rotation does not appear directly in the energy budget. The essence of the proof is that global rotation strongly reduces the energy transfers from the 2D base-flow to the 3D perturbations: for strong enough rotation, these transfers are too weak to overcome viscous damping, and the 3D perturbations decay in the long-time limit. As a result, the 2D base-flow is stable.

The fact that global rotation reduces the transfers between the 2D and the vertically dependent 3D modes has been known since \citet{Greenspan} and \citet{Waleffe}: in the asymptotic limit of low Rossby number, the 3D modes can be described in terms of weakly nonlinear inertial waves, and the dominant interaction between such waves consists of resonant triads. However, such resonant triads cannot transfer energy from the 2D modes to the 3D ones. This key result is obtained through a perturbative analysis and is valid to lowest order in Rossby number only. At higher order in Rossby number, near-resonant triads and four-wave interactions can transfer energy between the 2D and 3D modes \citep{SmithWaleffe}. 

In the field of mathematical analysis, similar results were obtained by \citet{Babin1, Babin2}, who translated into rigorous and exact analysis the concept of averaging over the fast rotation period, to study the regularity of solutions to the rotating Euler and Navier-Stokes equations. The proof of absolute two-dimensionalization that we present in section \ref{abstwodim} makes extensive use of their theorem on the existence of bounded solutions to the rapidly rotating 3D Navier-Stokes equation (theorem 1 in \citet{Babin2}).

Along the way to proving this theorem, they provide a decomposition of rotating flows on a time interval $[0, T]$ into three components: a 2D flow satisfying the 2D Navier-Stokes equation, some inertial waves that are advected and sheared by the 2D flow, and a small remainder. The wave part follows a reduced system of equations with coefficients depending on the 2D flow, and can be solved for exactly in some cases. The remainder decreases as $\Omega^{-1/2}$ but increases rapidly (typically exponentially) with the length $T$ of the time interval. This decomposition is useful when the remainder is indeed small, that is to say, in the limit of very fast rotation, for a given time interval. However, it cannot be used as is in the long-time limit ($T \to \infty$) to answer the central question raised above. Here we therefore address stability to 3D perturbations head-on. 

In the framework of geophysical fluid dynamics, the 2D base-flow corresponds to ``balanced" fluid motion in the 2D slow manifold, while instability with respect to 3D perturbations corresponds to spontaneous wave generation \citep{Vanneste}. The present study focuses on body forces that input energy directly into the 2D modes: we prove that the corresponding (possibly turbulent) flow settles exactly in the 2D slow manifold for low enough Rossby number, with no spontaneous wave generation.

We introduce the setup and notations in section \ref{2Dsol}, before describing the two-dimensional solutions to the three-dimensional problem. In sections \ref{linpert} to \ref{suffcond}, we consider the 3D rotating Navier-Stokes equation linearized about such a 2D base flow, and we prove the linear stability of these 2D solutions to 3D perturbations, for rapid global rotation: we derive sufficient criteria for such linear stability, either in terms of the Reynolds and Rossby numbers, or in terms of the Grashof and forcing-based Rossby numbers. The proof itself consists of 5 steps:
\begin{itemize}
\item Write the evolution equation for the kinetic energy of the 3D perturbation.
\item Introduce a cut-off wavenumber ${\cal K}$, and control the small scales of the perturbation (wave numbers larger than ${\cal K}$) with the viscous damping term (section \ref{linpert}).
\item Decompose the perturbation into helical modes (section \ref{helicalpart}).
\item Using an integration by parts in time, show that the transfer of energy from the 2D base flow to the large scales of the 3D perturbation (wave numbers smaller than ${\cal K}$) is inversely proportional to the rotation rate $\Omega$ (section \ref{controlover}).
\item For rapid global rotation, this transfer term is therefore weaker than the viscous damping of the 3D perturbation, hence the  stability criterion (section \ref{suffcond}). \\
\end{itemize} 

In section \ref{abstwodim}, we consider arbitrary initial conditions for the velocity field, with arbitrarily large vertically-dependent perturbations. We repeat the steps listed above to prove absolute two-dimensionalization for fast enough global rotation.

On first reading, one may want to go directly from the end of section \ref{2Dsol} to the concluding section \ref{discsection}, where we comment on the physical implications of these results.


%
%

\section{Rotating turbulence in a periodic domain and 2D solutions\label{2Dsol}}

\begin{figure}
\begin{center}
\includegraphics[width=120 mm]{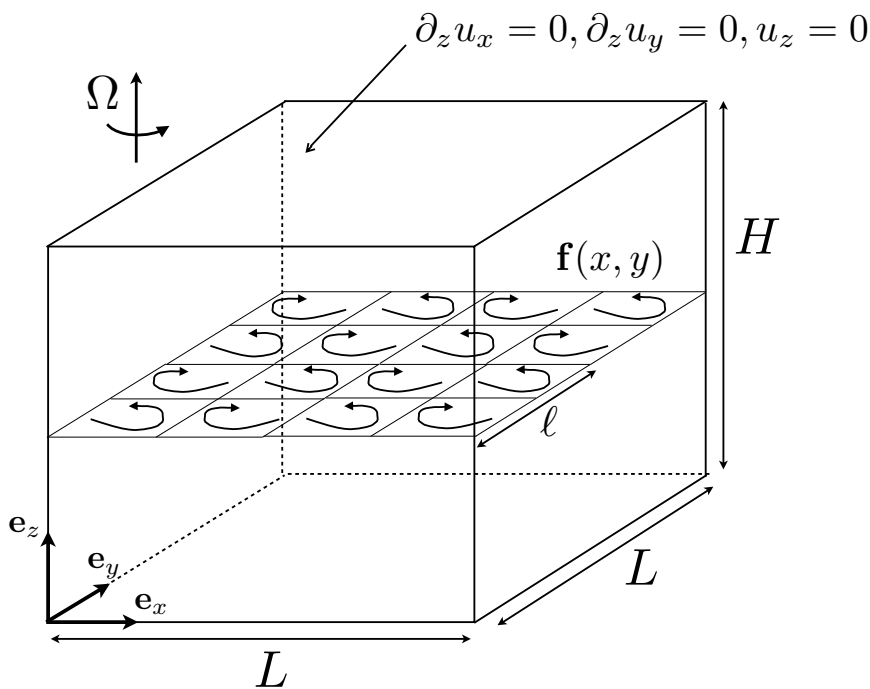}
\end{center}
\caption{Flow of a Newtonian fluid in a frame rotating at angular velocity $\Omega$ around the vertical $z$ axis. It is driven by a horizontal body-force $\textbf{f}$ that is independent of the vertical. We assume periodic boundary-conditions in the horizontal, and stress-free boundaries at $z=0$ and $z=H$.}
\label{3Dbox}
\end{figure}

\subsection{Body-forced rotating turbulence\label{bfrt}}

The setup is sketched in figure \ref{3Dbox}: an incompressible fluid of kinematic viscosity $\nu$ flows inside a domain $(x,y,z) \in \mathcal{D}=[0, L] \times [0,L] \times [0,H]$ with a Cartesian frame $(\textbf{e}_x,\textbf{e}_y,\textbf{e}_z)$. The fluid is subject to background rotation at a rate $\Omega$ around the $z$ axis, referred to as the vertical axis by convention. It is stirred by a steady divergence-free two-dimensional horizontal body-force $\textbf{f}(x,y)=(f_x,f_y,0)$ that is periodic on a scale $\ell$, an integer fraction of $L$. That is, ${\bf f}(x,y)=F {\boldsymbol \phi}(\frac{x}{\ell},\frac{y}{\ell})$, where ${\boldsymbol \phi}$  is periodic of period $1$ in each dimensionless variable, has vanishing spatial mean, and r.m.s. magnitude $1$. We refer to $F$ as the amplitude, and ${\boldsymbol \phi}$ as the shape of the force.  We consider periodic boundary conditions in the horizontal directions, and stress-free boundary conditions in the vertical (although the proofs of the present study easily carry over to a 3D periodic domain).
The velocity field ${\bf u}(x,y,z,t)$ follows the rotating Navier-Stokes equation,
\begin{equation}
\partial_t \textbf{u} +(\textbf{u} \cdot \bnabla) \textbf{u} + 2 \Omega {\bf e}_z \times \textbf{u} =-\bnabla p + \nu \Delta \textbf{u} + {\bf f} \, ,
\label{NS}
\end{equation}
together with the following boundary conditions at the top and and bottom boundaries,
\begin{equation}
\partial_z u_x=0\,, \partial_z u_y=0\,, u_z=0, \mbox{at } z=0 \mbox{ and } z=H \, .
\label{BC}
\end{equation}
We consider the solutions of equation (\ref{NS}) that have vanishing total momentum initially, and therefore at any subsequent time: the spatial average of the velocity field ${\bf u}$ is zero at all time.
From equation (\ref{NS}) we define the Reynolds number and the Rossby number based on the root-mean-square velocity $U$ of the flow, where the mean is performed over space and time:
\begin{equation}
Re=\frac{U \ell}{\nu} \, , \quad Ro=\frac{U}{\ell \Omega} \, .\label{nondimnumb}
\end{equation}

We stress the fact that these dimensionless numbers combine the root-mean-square velocity $U$ of the solution with the forcing scale $\ell$, instead of the typical scale of the velocity field (the integral scale). While the integral scale is probably close to $\ell$ for non-rotating 3D turbulence, it may increase very greatly and even reach the domain size $L$ for strong rotation, because of two-dimensionalization and enhanced inverse energy transfers. Nevertheless, the Reynolds and Rossby numbers defined in (\ref{nondimnumb}) are familiar to the theory of rotating turbulence, as well as to experimentalists: single-point velocity measurements usually lead to a good estimate of the root-mean square velocity and allow one to estimate $Re$ and $Ro$. Without loss of rigor, we will therefore present some results in terms of $Re$ and $Ro$ defined in (\ref{nondimnumb}).

We also introduce dimensionless numbers based on the strength $F$ of the forcing. The Grashof number and the forcing Rossby number are
\begin{equation}
Gr=\frac{F \ell^3}{\nu^2} \, , \quad Ro^{(f)}=\frac{\sqrt{F}}{\sqrt{\ell} \Omega} \, .\label{nondimnumbF}
\end{equation}
In contrast with $Re$ and $Ro$, $Gr$ and $Ro^{(f)}$ are control parameters: they do not require knowledge of the solution to be evaluated. For instance, $Gr$ and $Ro^{(f)}$ can be specified at the outset of a numerical simulation. In the following we present both results expressed in terms of $Re$ and $Ro$, which are useful for qualitative comparison with boundary-driven experiments, and results expressed with $Gr$ and $Ro^{(f)}$, which are useful for comparison with body-forced numerical simulations or  experiments.


In the following we use many inequalities. To alleviate the algebra somewhat, we make extensive use of the notation $\lesssim$, where $a \lesssim b$ means that there is a dimensionless constant $c>0$ such that $a \leq c b$, where the constant $c$ is independent of the parameters of the problem: $\nu$, $F$, $U$, $\ell$, $L$, $H$, etc. This constant can depend only on the precise choice of the dimensionless shape function $\boldsymbol \phi$ of the forcing. In the following, we denote as $c$ any such positive $\mathcal{O}(1)$ constant, and we sometimes use the same symbol $c$ to denote different constants in successive lines of algebra. Numbered constants $c_i$ ($\tilde{c}_i$ in the appendix) keep the same value between different lines of algebra.

Finally, we consider only domains that are cubic or shallower than a cube, $H \leq L$, and because we focus on the large-Reynolds-number behavior of the system, we restrict attention to $Re \geq 2$ and $Gr \geq 2$.

\subsection{Two-dimensional solutions}

Equation (\ref{NS}) with the boundary conditions (\ref{BC}) admits vertically invariant 2D solutions ${\bf u}(x,y,z,t)={\bf V}(x,y,t)$, where ${\bf V}$ satisfies the two-dimensional Navier-Stokes equation,
\begin{equation}
\partial_t \textbf{V} +(\textbf{V} \cdot \bnabla) \textbf{V}  =-\bnabla p + \nu \Delta \textbf{V} + {\bf f} \, .
\label{NS2D}
\end{equation}
Note that, in the 2D Navier-Stokes equation, the Coriolis force is a gradient that can be absorbed into the pressure term: $\Omega$ disappears from the equation (and we keep using $p$ to denote the modified pressure). ${\bf V}(x,y,t)$ is a horizontal velocity field, with vanishing vertical component (for periodic boundary conditions in the vertical, the vertical component of ${\bf V}$ satisfies a sourceless advection-diffusion equation and therefore vanishes in the long-time limit).


Rigorous bounds on the time-averaged enstrophy and enstrophy dissipation rate can be computed for solutions of the 2D Navier-Stokes equation (\ref{NS2D}). The derivation of these bounds is recalled in appendix \ref{AppendixA}. In terms of the forcing amplitude $F$, and denoting as $\omega$ the vertical vorticity of the 2D flow $\textbf{V}$, we obtain
\begin{eqnarray}
 \left< \| \omega \|_2^2 \right> \lesssim  \frac{F^2 \ell^2 L^2 H}{\nu^2} \, , \label{boundEnstrophyF} \\
  \left< \| \bnabla \omega \|_2^2 \right> \lesssim \frac{F^2 L^2 H}{\nu^2}  \, , \label{boundEDF} 
\end{eqnarray}
where $\la \dots \ra$ denotes time average, and $\| \dots \|_2$ is the standard $L_2$ norm in 3D:
\begin{eqnarray}
\| {\bf h} \|_2^2= \int_\mathcal{D} | {\bf h} |^2 \, \mathrm{d}^3 \textbf{x} \, .
\end{eqnarray}
Alternate bounds were obtained by \citet{AlexakisDoering} in terms of the r.m.s. velocity $U$. Using our notations, their equations (23) and (19) translate into
\begin{eqnarray}
 \left< \| \omega \|_2^2 \right> \lesssim  \frac{H L^2 U^2}{\ell^2} \sqrt{Re} \, , \label{boundEnstrophy} \\
  \left< \| \bnabla \omega \|_2^2 \right> \lesssim \frac{H L^2 U^2}{\ell^4} Re  \, , \label{boundED} 
\end{eqnarray}
where we restrict attention to $Re \geq 2$.

These bounds for 2D flows can be further reduced if the forcing contains a single wavenumber in Fourier space, i.e., if it is such that ${\bf f}$ is an eigenmode of the Laplacian operator (see \citet{Constantin} for a description of 2D turbulence driven by such forcing). These forcings are sometimes called ``single-mode", or Kolmogorov forcings. For such forcings the improved bounds on the enstrophy and enstrophy dissipation rate are
\begin{eqnarray}
 \left< \| \omega \|_2^2 \right> \lesssim  \frac{H L^2 U^2}{\ell^2} \, , \label{boundEnstrophyKolmogorov} \\
\left< \| \bnabla \omega \|_2^2 \right> \lesssim \frac{H L^2 U^2}{\ell^4} \, . \label{boundEDKolmogorov}
\end{eqnarray}

\section{Linear perturbation to the 2D solution\label{linpert}}

Consider a 2D solution ${\bf V}(x,y,t)$ lying on the attractor of the 2D Navier-Stokes equation (\ref{NS2D}). Our goal is to prove linear two-dimensionalization: we wish to show that, for strong enough global rotation $\Omega$, this solution is stable with respect to infinitesimal 3D perturbations.
We therefore consider the evolution of an infinitesimal perturbation ${\bf v}(x,y,z,t)$ to the two-dimensional flow ${\bf V}(x,y,t)$. We write ${\bf u}(x,y,z,t)={\bf V}(x,y,t)+{\bf v}(x,y,z,t)$, where ${\bf v} \ll {\bf V}$ is infinitesimal, and consider the linearized evolution equation for ${\bf v}$:
\begin{equation}
\partial_t \textbf{v} +(\textbf{V} \cdot \bnabla) \textbf{v} +(\textbf{v} \cdot \bnabla) \textbf{V} + 2 \Omega {\bf e}_z \times \textbf{v} =-\bnabla p' + \nu \Delta \textbf{v} \, ,
\label{eqv}
\end{equation}
where $p'$ is the pressure perturbation. Because the base-flow is independent of the vertical, different vertical Fourier modes of ${\bf v}$ evolve independently. We therefore consider a perturbation that has a single wavenumber $K > 0$ in the vertical. More precisely, we consider the following Fourier decomposition of the perturbation,
\begin{equation}
\textbf{v} (x,y,z,t) = \sum\limits_{{\bf k}\in S_{\text{3D}}} \textbf{v}_{\bf k}(t) \, e^{i {\bf k} \cdot {\bf x}} \, ,
\label{Fourierv}
\end{equation}
where the wave vector ${\bf k}$ takes the values $(k_x,k_y,k_z)=(n_x \frac{2\pi}{L}, n_y \frac{2\pi}{L}, \pm K)$, with $(n_x,n_y) \in \mathbb{Z}^2$. We denote this set of wave vectors as $S_{\text{3D}}$.

Dotting $\textbf{v}$ into (\ref{eqv}) and integrating over the domain leads to the evolution equation for the $L_2$ norm of the perturbation:
\begin{equation}
\mathrm{d}_t \left( \frac{1}{2} \|\textbf{v}\|_2^2 \right) =  - \int_\mathcal{D} \textbf{v} \cdot (\bnabla \textbf{V}) \cdot \textbf{v} \, \mathrm{d}^3 \textbf{x} - \nu \|\bnabla \textbf{v}\|_2^2 \, ,
\label{eqv2}
\end{equation}
which we divide by $\|\textbf{v}\|_2^2$ to obtain
\begin{equation}
\frac{1}{2} \mathrm{d}_t \left( \ln  \|\textbf{v}\|_2^2 \right) =   \frac{- \int_\mathcal{D} \textbf{v} \cdot (\bnabla \textbf{V}) \cdot \textbf{v} \, \mathrm{d}^3 \textbf{x} - \nu \|\bnabla \textbf{v}\|_2^2}{\|\textbf{v}\|_2^2} \, .
\label{eqlnv}
\end{equation}
Our goal is to prove that, for large-enough $\Omega$, the time average of the right-hand-side of (\ref{eqlnv}) is negative, and therefore $\|\textbf{v}\|_2$ decays to zero in the long-time limit.

\subsection{Large versus small horizontal scales of the perturbation \label{largevssmall}}
From the spatial Fourier transform (\ref{Fourierv}) of ${\bf v}$, we define a cut-off ${\cal K}$ for the  wavenumber $k=|{\bf k}|$, and we write ${\bf v}={\bf v}_< + {\bf v}_>$, where ${\bf v}_<$ contains all the Fourier modes with $k \leq {\cal K}$ and ${\bf v}_>$ contains all the Fourier modes with $k > {\cal K}$. Equation (\ref{eqlnv}) becomes
\begin{eqnarray}
\frac{1}{2} \mathrm{d}_t \left( \ln  \|\textbf{v}\|_2^2 \right) & = &   \frac{1}{\|\textbf{v}\|_2^2} \left[- \int_\mathcal{D} \left[ \textbf{v}_< \cdot (\bnabla \textbf{V}) \cdot \textbf{v}_< +\textbf{v}_< \cdot (\bnabla \textbf{V}) \cdot \textbf{v}_> +\textbf{v}_> \cdot (\bnabla \textbf{V}) \cdot \textbf{v}_<  \right. \right. \label{eqlnvdecomp} \\
\nonumber  & & + \left. \left. \textbf{v}_> \cdot (\bnabla \textbf{V}) \cdot \textbf{v}_>  \right]\, \mathrm{d}^3 \textbf{x}  - \nu \|\bnabla \textbf{v}\|_2^2  \right] \, ,
\end{eqnarray}
We now bound all the terms on the right-hand side that involve $\textbf{v}_>$. Using successively H\"older's, the Cauchy-Schwarz, and Young's inequalities together with an optimization,
\begin{eqnarray}
& & \left| \int_\mathcal{D}  \left[ \textbf{v}_< \cdot (\bnabla \textbf{V}) \cdot \textbf{v}_> +\textbf{v}_> \cdot (\bnabla \textbf{V}) \cdot \textbf{v}_<   +  \textbf{v}_> \cdot (\bnabla \textbf{V}) \cdot \textbf{v}_>  \right] \, \mathrm{d}^3 \textbf{x} \right| \label{sumtriple}\\
\nonumber & & \leq \| \bnabla {\bf V} \|_\infty \left[ 2 \| \textbf{v}_<\|_2 \| \textbf{v}_>\|_2 +\| \textbf{v}_>\|_2^2 \right] \\
\nonumber & & \leq \frac{\nu {\cal K}^2}{4} \| \textbf{v}_>\|_2^2 + \frac{4}{\nu {\cal K}^2} \| \bnabla {\bf V} \|_\infty^2 \| \textbf{v}_<\|_2^2 + \frac{\nu {\cal K}^2}{4} \| \textbf{v}_>\|_2^2 + \frac{1}{\nu {\cal K}^2} \| \bnabla {\bf V} \|_\infty^2 \| \textbf{v}_>\|_2^2 \\
\nonumber & & \leq \frac{\nu {\cal K}^2}{2} \| \textbf{v}_>\|_2^2 + \frac{4}{\nu {\cal K}^2} \| \bnabla {\bf V} \|_\infty^2 \| \textbf{v}\|_2^2 \, ,
\end{eqnarray}
where $\| \dots \|_\infty$ denotes the standard $L_{\infty}$ norm in space. From Poincar\'e's inequality, $\|\bnabla \textbf{v}_>\|_2^2 \geq {\cal K}^2 \| \textbf{v}_> \|_2^2$ and $\|\bnabla \textbf{v}\|_2^2 \geq  \frac{\pi^2}{H^2}  \| \textbf{v} \|_2^2$, hence
\begin{eqnarray}
-\nu \|\bnabla \textbf{v}\|_2^2 \leq -\frac{\nu}{4} \|\bnabla \textbf{v}\|_2^2 - \frac{\nu {\cal K}^2}{2}  \| \textbf{v}_> \|_2^2 -\frac{\nu \pi^2}{4 H^2}  \| \textbf{v} \|_2^2 \, .
\label{decompvisc}
\end{eqnarray}
Inserting inequalities (\ref{sumtriple}) and (\ref{decompvisc}) in (\ref{eqlnvdecomp}) results in
\begin{eqnarray}
\frac{1}{2} \mathrm{d}_t \left( \ln  \|\textbf{v}\|_2^2 \right) & \leq &   \frac{1}{\|\textbf{v}\|_2^2} \left[- \int_\mathcal{D} \textbf{v}_< \cdot (\bnabla \textbf{V}) \cdot \textbf{v}_<  \, \mathrm{d}^3 \textbf{x}  - \frac{\nu}{4} \|\bnabla \textbf{v}\|_2^2 \right] +    \lambda_1(t) \, , \label{ineqlnvLS}
\end{eqnarray}
where $\lambda_1(t)  =    \frac{4}{\nu {\cal K}^2} \| \bnabla {\bf V} \|_\infty^2 - \frac{\nu \pi^2}{4 H^2}$.
Notice that the triple velocity product in (\ref{ineqlnvLS}) does not involve ${\bf v}_>$ anymore. In the following we choose the value 
\begin{eqnarray}
{\cal K}=\frac{\sqrt{32} H}{\pi \nu}  \sqrt{ \la \| \bnabla {\bf V} \|_\infty^2 \ra} \, ,\label{valueK}
\end{eqnarray}
which leads to the following time-average of $\lambda_1(t)$,
\begin{eqnarray}
\la \lambda_1 \ra = - \frac{\nu \pi^2}{8 H^2} \label{boundtalambda1}
\end{eqnarray}
which is sufficient control over small horizontal scales ${\bf v}_>$ of the perturbation. The quantity $\la \| \bnabla {\bf V} \|_\infty^2 \ra$ is bounded from above in appendix \ref{AppendixA}, which provides an upper bound on ${\cal K}$,
\begin{eqnarray}
{\cal K} \lesssim \frac{\sqrt{H}}{\nu} \sqrt{ \la \| \bnabla \omega \|_2^2 \ra } \ln^{1/2} \left( Gr \frac{L}{\ell} \right) \, ,\label{boundK}
\end{eqnarray}
where the right-hand-side involves the time-averaged enstrophy dissipation rate of the 2D base flow, which can be bounded in terms of its r.m.s. velocity $U$ using (\ref{boundED}) or (\ref{boundEDKolmogorov}), or in terms of the forcing strength $F$ using (\ref{boundEDF}).

\section{Helical wave decomposition\label{helicalpart}}

We perform a standard helical wave decomposition of the Fourier amplitudes of the velocity perturbation \citep{CambonJacquin, Waleffe},
\begin{eqnarray}
{\bf v}_{\bf k}(t)=b_+({\bf k},t) e^{ i \sigma_+({\bf k}) t}\, {\bf h}_+({\bf k})+ b_-({\bf k},t) e^{ i \sigma_-({\bf k}) t}\, {\bf h}_-({\bf k}) \, ,\label{helical}
\end{eqnarray}
%
where $\sigma_{s_{\bf k}}({\bf k})=s_{\bf k} 2 \Omega  k_z/k$ is the frequency of a linear inertial wave with spatial structure ${\bf h}_{s_{\bf k}}({\bf k})$. For non-vertical wave vectors, the latter is given by
\begin{equation}
{\bf h}_{s_{\bf k}}({\bf k})  =  \frac{1}{\sqrt{2}} \left[  \frac{({\bf e}_z \times {\bf e}_{\bf k})}{ |{\bf e}_z \times {\bf e}_{\bf k}|} \times {\bf e}_{\bf k} + i s_{\bf k} \frac{({\bf e}_z \times {\bf e}_{\bf k})}{ |{\bf e}_z \times {\bf e}_{\bf k}|} \right]  \, ,
\end{equation}
with ${\bf e}_{\bf k}$ the unit vector along ${\bf k}$, and $s_{\bf k}=\pm$ a sign coefficient. For vertical wave vectors, the structure of the helical modes is ${\bf h}_{s_{\bf k}}({\bf k})= \frac{1}{\sqrt{2}}({\bf e}_x + i s_{\bf k} \frac{k_z}{k}{\bf e}_y)$. These vectors are parallel to their curl, $i {\bf k} \times {\bf h}_{s_{\bf k}} = s_{\bf k} k {\bf h}_{s_{\bf k}}$, and they are normalized, $|{\bf h}_{s_{\bf k}}|=1$.

To obtain a similar decomposition for the $z$-independent base-flow ${\bf V}$, we first write it as a Fourier series,
\begin{equation}
\textbf{V} (x,y,t) = \sum\limits_{{\bf k}\in S_{\text{2D}}} \textbf{V}_{\bf k}(t) \, e^{i {\bf k} \cdot {\bf x}} \, ,
\label{FourierV}
\end{equation}
where the set $S_{\text{2D}}$ contains all the horizontal wave vectors of the periodic domain: 
${\bf k}=(k_x,k_y,k_z)=(n_x \frac{2\pi}{L}, n_y \frac{2\pi}{L}, 0)$, with $(n_x,n_y) \in \mathbb{Z}^2 \setminus\{ (0,0) \}$ (recall that $\textbf{V}$ has a vanishing average over the domain, see section \ref{bfrt}).
Because the frequency $\sigma_{s_{\bf k}}({\bf k})$ vanishes for horizontal wave vectors, the helical decomposition of each Fourier amplitude ${\bf V}_{\bf k}$ yields simply
\begin{eqnarray}
{\bf V}_{\bf k}=B_+({\bf k},t) \, {\bf h}_+({\bf k})+ B_-({\bf k},t) \, {\bf h}_-({\bf k}) \, ,\label{helicalV}
\end{eqnarray}
and because ${\bf V}$ is a horizontal flow, we get the additional relations 
\begin{eqnarray}
& & B_+({\bf k},t) + B_-({\bf k},t) = 0 \, , \\ \label{horizontalV}
& & |{\bf V}_{\bf k}|^2=2 |B_+({\bf k},t)|^2 = 2 |B_-({\bf k},t)|^2 \, .
\end{eqnarray}

For brevity, in the following we often write $b_{s_{\bf k}}$, $B_{s_{\bf k}}$ and $\sigma_{s_{\bf k}}$ only to designate respectively $b_{s_{\bf k}}({\bf k},t)$, $B_{s_{\bf k}}({\bf k},t)$ and $\sigma_{s_{\bf k}}({\bf k})$.

The oscillatory phases in the decomposition (\ref{helical}) absorb the Coriolis force: inserting the decompositions (\ref{Fourierv}), (\ref{helical}), (\ref{FourierV}) and (\ref{helicalV}) into the curl of (\ref{eqv}), we obtain
\begin{eqnarray}
\partial_t b_{s_{\bf k}}= -\nu k^2 b_{s_{\bf k}} + \sum\limits_{\substack{{\bf p}+{\bf q}+{\bf k}=0 \\ s_{\bf p}; s_{\bf q} }}  b_{s_{\bf p}}^* B_{s_{\bf q}}^* e^{-i(\sigma_{s_{\bf p}}+\sigma_{s_{\bf k}})t} \, C_{{\bf k} {\bf p} {\bf q}}^{s_{\bf k} s_{\bf p} s_{\bf q}} \, ,\label{dtb}
\end{eqnarray}
where the sum is over all ${\bf p}\in S_{\text{3D}}$ and ${\bf q}\in S_{\text{2D}}$ such that ${\bf p}+{\bf q}+{\bf k}=0$ and over the two sign coefficients $s_{\bf p}=\pm 1$ and $s_{\bf q}=\pm 1$. The following algebra involves many similar sums over wave vectors. In such sums, we omit to mention the sets in which the wave vectors are. In section \ref{controlover}, a sum over ${\bf k}$ involving a base-flow component $B_{s_{\bf k}}$ implies ${\bf k} \in S_{\text{2D}}$, while a sum over ${\bf k}$ involving a perturbation component $b_{s_{\bf k}}$ implies ${\bf k} \in S_{\text{3D}}$. We only mention the additional constraints under the sum symbol, e.g., ${\bf p}+{\bf q}+{\bf k}=0$. In section \ref{abstwodim}, we consider perturbations of arbitrary amplitude, and the sums over ${\bf k}$ involving a perturbation component $b_{s_{\bf k}}$ are over wave vectors ${\bf k}=(k_x,k_y,k_z)=(n_x \frac{2\pi}{L}, n_y \frac{2\pi}{L}, n_z \frac{\pi}{H})$, with $(n_x,n_y,n_z) \in \mathbb{Z}\times\mathbb{Z} \times (\mathbb{Z}  \setminus\{0 \} )$, unless otherwise stated under the sum sign.

The coupling coefficient in (\ref{dtb}) is 
\begin{equation}
C_{{\bf k} {\bf p} {\bf q}}^{s_{\bf k} s_{\bf p} s_{\bf q}}=2 (s_{\bf p} p- s_{\bf q} q ) g_{{\bf k} {\bf p} {\bf q}}^{s_{\bf k} s_{\bf p} s_{\bf q}} \, ,\label{defCkpq}
\end{equation}
with 
\begin{equation}
g_{{\bf k} {\bf p} {\bf q}}^{s_{\bf k} s_{\bf p} s_{\bf q}}= \frac{1}{2} ({\bf h}_{s_{\bf k}}^*\times {\bf h}_{s_{\bf q}}^*)\cdot {\bf h}_{s_{\bf p}}^* \, .\label{defg}
\end{equation}

Using the decompositions (\ref{helical}) and (\ref{helicalV}), the triple velocity product in (\ref{ineqlnvLS}) reads
\begin{eqnarray}
\int_\mathcal{D} \textbf{v}_< \cdot (\bnabla \textbf{V}) \cdot \textbf{v}_<  \, \mathrm{d}^3 \textbf{x} = L^2 H \sum\limits_{\substack{{\bf p}+{\bf q}+{\bf k}=0 \\ k \leq {\cal K}; p \leq {\cal K} \\ s_{\bf p}; s_{\bf q}; s_{\bf k} }} B_{s_{\bf q}} b_{s_{\bf k}} b_{s_{\bf p}} s_{\bf k} k e^{i(\sigma_{s_{\bf k}}+\sigma_{s_{\bf p}})t} 2 g_{ {\bf q} {\bf k} {\bf p}}^{*  s_{\bf q} s_{\bf k} s_{\bf p}} \, ,
\end{eqnarray}
or, because ${\bf k}$ and ${\bf p}$ have symmetrical roles,
\begin{eqnarray}
\int_\mathcal{D} \textbf{v}_< \cdot (\bnabla \textbf{V}) \cdot \textbf{v}_<  \, \mathrm{d}^3 \textbf{x} = L^2 H \sum\limits_{\substack{{\bf p}+{\bf q}+{\bf k}=0 \\ k \leq {\cal K}; p \leq {\cal K}\\ s_{\bf p}; s_{\bf q}; s_{\bf k} }} B_{s_{\bf q}} b_{s_{\bf k}} b_{s_{\bf p}}  e^{i(\sigma_{s_{\bf k}}+\sigma_{s_{\bf p}})t} (s_{\bf k} k-s_{\bf p} p) g_{ {\bf q} {\bf k} {\bf p}}^{*  s_{\bf q} s_{\bf k} s_{\bf p}} \, . \qquad
\end{eqnarray}
Because the sum is over ${\bf p}+{\bf q}+{\bf k}=0$, $p_z=-k_z$, and we can write
\begin{eqnarray}
\int_\mathcal{D} \textbf{v}_< \cdot (\bnabla \textbf{V}) \cdot \textbf{v}_<  \, \mathrm{d}^3 \textbf{x} = - L^2 H \sum\limits_{\substack{{\bf p}+{\bf q}+{\bf k}=0 \\ k \leq {\cal K}; p \leq {\cal K}\\ s_{\bf p}; s_{\bf q}; s_{\bf k} }} B_{s_{\bf q}} b_{s_{\bf k}} b_{s_{\bf p}}  e^{i(\sigma_{s_{\bf k}}+\sigma_{s_{\bf p}})t} s_{\bf k} s_{\bf p} k p \frac{\sigma_{s_{\bf k}}+\sigma_{s_{\bf p}}}{2 \Omega k_z} g_{ {\bf q} {\bf k} {\bf p}}^{*  s_{\bf q} s_{\bf k} s_{\bf p}} \, . \qquad \label{tripleproduct}
\end{eqnarray}
We see in this equation the result of \citet{Waleffe} and \citet{Greenspan}: the coupling coefficient between the 2D base-flow and a couple of inertial waves is proportional to $\sigma_{s_{\bf k}}+\sigma_{s_{\bf p}}$, and therefore it vanishes for resonant triads, i.e., for $\sigma_{s_{\bf k}}+\sigma_{s_{\bf p}}=0$.

\section{Control over the large scales of the perturbation\label{controlover}}

Integrate (\ref{ineqlnvLS}) from time $t=0$ to $t=T$ and divide by $T$ to obtain
\begin{eqnarray}
 \frac{1}{T} \ln  \frac{  \|\textbf{v}\|_2(t=T)}{ \|\textbf{v}\|_2(t=0)} & \leq & - \frac{1}{T }\int_{t=0}^{t=T} \frac{\int_\mathcal{D} \textbf{v}_< \cdot (\bnabla \textbf{V}) \cdot \textbf{v}_<  \, \mathrm{d}^3 \textbf{x}}{\| {\bf v} \|_2^2} \mathrm{d}t  \label{integrateineq}\\
\nonumber & &   - \frac{\nu}{4T} \int_{t=0}^{t=T} \frac{ \|\bnabla \textbf{v}\|_2^2}{\|\textbf{v}\|_2^2}\mathrm{d}t   +    \frac{1}{T} \int_{t=0}^{t=T} \lambda_1(t) \mathrm{d}t \, . 
\end{eqnarray}

Our goal is to show that the right-hand-side of this inequality is negative in the long-time-$T$ limit provided that $\Omega$ is large enough: this ensures that $\|\textbf{v}\|_2$ decays. The key step is to prove that the contribution from the triple velocity product is small when $\Omega$ is large. To wit, we insert the expression (\ref{tripleproduct}) of the triple velocity product, before performing an integration by parts in time, where we integrate the oscillatory exponential (and differentiate the rest of the integrand). This gives
\begin{eqnarray}
 - \frac{1}{T }\int_{t=0}^{t=T} \frac{\int_\mathcal{D} \textbf{v}_< \cdot (\bnabla \textbf{V}) \cdot \textbf{v}_<  \, \mathrm{d}^3 \textbf{x}}{\| {\bf v} \|_2^2} \mathrm{d}t & = & \frac{L^2 H}{2 \Omega k_z}( {\cal T}_1 + {\cal T}_2 +{\cal T}_3 + {\cal T}_4 )\, ,\label{fourterms}
\end{eqnarray}
where 
\begin{eqnarray}
{\cal T}_1 & = & \frac{1}{T } \int_{t=0}^{t=T} \left[ \sum\limits_{\substack{{\bf p}+{\bf q}+{\bf k}=0 \\ k \leq {\cal K}; p \leq {\cal K}; \sigma_{s_{\bf k}}+\sigma_{s_{\bf p}} \neq 0\\ s_{\bf p}; s_{\bf q}; s_{\bf k} }} \partial_t(B_{s_{\bf q}})   \frac{b_{s_{\bf k}} b_{s_{\bf p}}}{\| {\bf v} \|_2^2}  e^{i(\sigma_{s_{\bf k}}+\sigma_{s_{\bf p}})t}  i s_{\bf k} s_{\bf p} k p \, {g}_{ {\bf q} {\bf k} {\bf p}}^{*   s_{\bf q} s_{\bf k} s_{\bf p}}   \right] \mathrm{d}t \label{defT1}\\
{\cal T}_2 & = & \frac{1}{T } \int_{t=0}^{t=T} \left[ \sum\limits_{\substack{{\bf p}+{\bf q}+{\bf k}=0 \\ k \leq {\cal K}; p \leq {\cal K}; \sigma_{s_{\bf k}}+\sigma_{s_{\bf p}} \neq 0 \\ s_{\bf p}; s_{\bf q}; s_{\bf k} }} B_{s_{\bf q}}   \frac{b_{s_{\bf p}} \partial_t b_{s_{\bf k}}  + b_{s_{\bf k}} \partial_t b_{s_{\bf p}} }{\| {\bf v} \|_2^2}  e^{i(\sigma_{s_{\bf k}}+\sigma_{s_{\bf p}})t}  i s_{\bf k} s_{\bf p} k p \, {g}_{ {\bf q} {\bf k} {\bf p}}^{*   s_{\bf q} s_{\bf k} s_{\bf p}}   \right] \mathrm{d}t \qquad \\
{\cal T}_3 & = & \frac{1}{T } \int_{t=0}^{t=T} \left[ \sum\limits_{\substack{{\bf p}+{\bf q}+{\bf k}=0 \\ k \leq {\cal K}; p \leq {\cal K} ; \sigma_{s_{\bf k}}+\sigma_{s_{\bf p}} \neq 0\\ s_{\bf p}; s_{\bf q}; s_{\bf k} }} B_{s_{\bf q}}   b_{s_{\bf p}}  b_{s_{\bf k}}   e^{i(\sigma_{s_{\bf k}}+\sigma_{s_{\bf p}})t}  i s_{\bf k} s_{\bf p} k p \, {g}_{ {\bf q} {\bf k} {\bf p}}^{*   s_{\bf q} s_{\bf k} s_{\bf p}}  \frac{-\mathrm{d}_t({\| {\bf v} \|_2^2})}{\| {\bf v} \|_2^4} \right] \mathrm{d}t \\
{\cal T}_4 & = &\frac{1}{T} \left[ \sum\limits_{\substack{{\bf p}+{\bf q}+{\bf k}=0 \\ k \leq {\cal K}; p \leq {\cal K} ; \sigma_{s_{\bf k}}+\sigma_{s_{\bf p}} \neq 0\\ s_{\bf p}; s_{\bf q}; s_{\bf k} }} B_{s_{\bf q}} \frac{b_{s_{\bf k}} b_{s_{\bf p}}}{\| {\bf v} \|_2^2}  e^{i(\sigma_{s_{\bf k}}+\sigma_{s_{\bf p}})t}  (-i) s_{\bf k} s_{\bf p} k p \, {g}_{ {\bf q} {\bf k} {\bf p}}^{*   s_{\bf q} s_{\bf k} s_{\bf p}}  \right]_{t=0}^{t=T} .\label{defT4}
\end{eqnarray}
To bound these terms, we use two inequalities on the coupling coefficients:
\begin{itemize}
\item Because the helical vectors ${\bf h}_{s_{\bf k}}$ are normalized, $|{g}_{ {\bf q} {\bf k} {\bf p}}^{  s_{\bf q} s_{\bf k} s_{\bf p}}| \lesssim 1$. For wave vectors such that $k \leq {\cal K}$, and $p \leq {\cal K}$, we therefore have
\begin{equation}
|k p \, {g}_{ {\bf q} {\bf k} {\bf p}}^{  s_{\bf q} s_{\bf k} s_{\bf p}} | \lesssim {\cal K}^2 \, . \label{ineqkpg}
\end{equation}
\item Using $|{g}_{ {\bf q} {\bf k} {\bf p}}^{  s_{\bf q} s_{\bf k} s_{\bf p}}| \lesssim 1$ we also obtain
\begin{equation}
| {C}_{{\bf k} {\bf p} {\bf q}}^{s_{\bf k} s_{\bf p} s_{\bf q}} | \lesssim p+q \label{ineqC} \, .
\end{equation}
\end{itemize}
Because the domain has a finite vertical extent $H$, the vertical wavenumber satisfies $K \geq \frac{\pi}{H}$ for stress-free top and bottom boundaries (and $K \geq \frac{2 \pi}{H}$ for periodic boundary conditions). In either case, $K \gtrsim 1/H$. Let us make use of (\ref{ineqkpg}) and the Cauchy-Schwarz inequality to bound ${\cal T}_1$:
\begin{eqnarray}
|{\cal T}_1| & \lesssim & \frac{{\cal K}^2}{ T } \int_{t=0}^{t=T} \left[ \sum\limits_{\substack{{\bf p}+{\bf q}+{\bf k}=0 \\ k \leq {\cal K}; p \leq {\cal K}\\ s_{\bf p}; {s_{\bf q}}; s_{\bf k} }}  |\partial_t(B_{s_{\bf q}})|   \frac{|b_{s_{\bf k}}| | b_{s_{\bf p}}|}{\| {\bf v} \|_2^2}    \right] \mathrm{d}t \\
\nonumber & \lesssim & \frac{{\cal K}^2}{L^2 H T} \int_{t=0}^{t=T} \left[ \sum\limits_{\substack{ {\bf q} , q \leq 2 {\cal K}  \\ s_{\bf q} }}  |\partial_t(B_{s_{\bf q}}({\bf q}))|     \right] \mathrm{d}t \, .
\end{eqnarray}
Taking the limit $T\to \infty$,
\begin{eqnarray}
\lim_{T \to \infty} |{\cal T}_1| & \lesssim & \frac{{\cal K}^2}{L^2 H} \left< \sum\limits_{\substack{ {\bf q} , q \leq 2 {\cal K}  \\ s_{\bf q} }} |\partial_t(B_{s_{\bf q}})|     \right> \, .
\end{eqnarray}
The time-averaged quantity appearing in the right-hand side is bounded in appendix \ref{AppendixA} in terms of the time-averaged energy and enstrophy dissipation rates inside the 2D base flow (see \ref{bounddt}). The resulting bound on $\lim_{T \to \infty} |{\cal T}_1|$ is
\begin{eqnarray}
\nonumber \lim_{T \to \infty} |{\cal T}_1| & \lesssim & \frac{{\cal K}^2}{L^2 H}  \left[ \frac{F L^2 }{\ell^2} +  \left(\frac{1}{H} \sqrt{\la \|  \omega \|_2^2 \ra \la \| \bnabla \omega \|_2^2 \ra} + \la \| \bnabla \omega \|_2^2 \ra \right) \ln \left( Gr  \frac{L}{\ell} \right) \right]\, ,\\ 
\end{eqnarray}
and after substituting the bound (\ref{boundK}) on ${\cal K}$,
\begin{eqnarray}
\nonumber \lim_{T \to \infty} |{\cal T}_1| & \lesssim &  \frac{1}{\nu^2 L^2 } \la \| \bnabla \omega \|_2^2 \ra  \ln \left( Gr  \frac{L}{\ell} \right) \\
& \times & \left[ \frac{F L^2}{\ell^2} +  \left(\frac{1}{H} \sqrt{\la \|  \omega \|_2^2 \ra \la \| \bnabla \omega \|_2^2 \ra} + \la \| \bnabla \omega \|_2^2 \ra \right) \ln \left( Gr  \frac{L}{\ell} \right) \right]\, , \quad
\label{limT1}
\end{eqnarray}

We now move on to term ${\cal T}_2$. Replacing the time-derivative of $b_{s_{\bf k}}$ by its expression (\ref{dtb}), and $p$ by $|{\bf k}+{\bf q} |$,
\begin{eqnarray}
\nonumber |{\cal T}_2| & \lesssim &  \frac{1}{T } \int_{t=0}^{t=T}    \left[ \frac{{\cal K}}{\| {\bf v} \|_2^2}   \sum\limits_{\substack{ {\bf q} ; q \leq 2 {\cal K}  \\ s_{\bf q} }}  |B_{s_{\bf q}}({\bf q})| \right. \\
\nonumber & & \left.  \sum\limits_{\substack{{\bf k} , k\leq {\cal K}, |{\bf k} + {\bf q}|\leq {\cal K}  \\ s_{\bf p}; s_{\bf k}}} | {\bf k}+{\bf q} |  |b_{s_{\bf p}}(-{\bf k}-{\bf q})|  \left| -\nu k^2  b_{s_{\bf k}}  +  \sum\limits_{\substack{{\bf m}+{\bf l}+{\bf k}=0 \\ s_{\bf m}; s_{\bf l} }} b_{s_{\bf l}}^* B_{s_{\bf m}}^* e^{-i(\sigma_{s_{\bf l}}+\sigma_{s_{\bf k}})t} {C}_{{\bf k} {\bf l} {\bf m}}^{s_{\bf k} s_{\bf l} s_{\bf m}}   \right|  \right] \mathrm{d}t \\
\nonumber & \lesssim & \frac{1}{T } \int_{t=0}^{t=T} \left[\frac{\nu {\cal K}^2}{  \| {\bf v} \|_2^2   } \sum\limits_{\substack{ {{\bf q}, q\leq 2{\cal K}} \\ s_{\bf q} }} |B_{s_{\bf q}}|  \sum\limits_{\substack{{\bf k}, k\leq {\cal K}, |{\bf k} + {\bf q}|\leq {\cal K} \\ s_{\bf p};s_{\bf k}}} |{\bf k}+{\bf q}| |b_{s_{\bf p}}(-{\bf k}-{\bf q})| k |b_{s_{\bf k}}| \right] \mathrm{d}t \\
\nonumber  &+&  \frac{{\cal K} }{ T}  \int_{t=0}^{t=T} \left[\frac{1}{\| {\bf v} \|_2^2} \sum\limits_{\substack{ {{\bf q}, q\leq 2{\cal K}} \\ s_{\bf q} }}  |B_{s_{\bf q}}|  \sum\limits_{\substack{{\bf k}, k\leq {\cal K}, |{\bf k} + {\bf q}|\leq {\cal K} \\ s_{\bf p}}} |{\bf k}+{\bf q}| |b_{s_{\bf p}}(-{\bf k}-{\bf q})|    \sum\limits_{\substack{{\bf m}+{\bf l}+{\bf k}=0 \\ s_{\bf m}; s_{\bf l}; s_{\bf k} }} |b_{s_{\bf l}}| |B_{s_{\bf m}}|   | {C}_{{\bf k} {\bf l} {\bf m}}^{s_{\bf k} s_{\bf l} s_{\bf m}} |  \right] \mathrm{d}t \, .\\
\label{ineqT2temp}
\end{eqnarray}
Using inequality (\ref{ineqC}) and the Cauchy-Schwarz inequality,
\begin{eqnarray}
 \sum\limits_{\substack{{\bf m}+{\bf l}+{\bf k}=0 \\ s_{\bf m}; s_{\bf l}; s_{\bf k} }} |b_{s_{\bf l}}| |B_{s_{\bf m}}|   | {C}_{{\bf k} {\bf l} {\bf m}}^{s_{\bf k} s_{\bf l} s_{\bf m}} | & \lesssim &  \sum\limits_{\substack{{\bf m}+{\bf l}+{\bf k}=0 \\ s_{\bf m};s_{\bf l}; s_{\bf k} }} |b_{s_{\bf l}}| |B_{s_{\bf m}}|   (l + m) \label{ineqtemp1}\\
  \nonumber & \lesssim & \sqrt{\sum\limits_{\substack{{\bf l}; s_{\bf l}}} l^2 |b_{s_{\bf l}}|^2  }  \sqrt{\sum\limits_{{\bf m}; s_{\bf m}} |B_{s_{\bf m}}|^2} + \sqrt{\sum\limits_{\substack{{\bf l}; s_{\bf l}}} |b_{s_{\bf l}}|^2  }  \sqrt{\sum\limits_{{\bf m};s_{\bf m}} m^2 |B_{s_{\bf m}}|^2} \\
 \nonumber & \lesssim & \frac{ \| {\bnabla \bf v} \|_2 \| {\bf V} \|_2 + \| {\bf v} \|_2  \| \omega \|_2 }{L^2 H} \, ,
\end{eqnarray}
and using the Cauchy-Schwarz inequality again,
\begin{eqnarray}
\nonumber \sum\limits_{\substack{{\bf k}, k\leq {\cal K}, |{\bf k} + {\bf q}|\leq {\cal K} \\ s_{\bf p}}} |{\bf k}+{\bf q}| |b_{s_{\bf p}}(-{\bf k}-{\bf q})| & \lesssim &  \sqrt{ \sum\limits_{\substack{{\bf k}, k\leq {\cal K}; s_{\bf p}}} |{\bf k}+{\bf q}|^2 |b_{s_{\bf p}}(-{\bf k}-{\bf q})|^2   }     \sqrt{ \sum\limits_{\substack{{\bf k}, k\leq {\cal K}}} 1 } \quad \\ 
 & \lesssim & \frac{{\cal K}  \| \bnabla {\bf v}_< \|_2}{\sqrt{H}} \, . \label{ineqtemp2}
\end{eqnarray}
The sum over $q$ is bounded using the Cauchy-Schwarz inequality once again,
\begin{eqnarray}
\sum\limits_{\substack{{\bf q}, q\leq 2{\cal K} \\ s_{\bf q}}} |B_{s_{\bf q}}|  \leq \sqrt{\sum\limits_{\substack{{\bf q}, q\leq 2{\cal K} \\ s_{\bf q}}} q^2 |B_{s_{\bf q}}|^2} \sqrt{\sum\limits_{\substack{{\bf q}, q\leq 2{\cal K} \\ s_{\bf q}}} \frac{1}{q^2}}     \lesssim {\| \omega \|_2} \sqrt{\frac{\ln(2 {\cal K}L)}{H}} \, .\label{ineqtemp3}
\end{eqnarray}
Using successively the inequalities (\ref{ineqtemp1}), (\ref{ineqtemp2}) and (\ref{ineqtemp3}), we bound the last time integral in (\ref{ineqT2temp}) according to
\begin{eqnarray}
\nonumber & & \frac{{\cal K} }{ T}  \int_{t=0}^{t=T} \left[\frac{1}{\| {\bf v} \|_2^2} \sum\limits_{\substack{{\bf q}, q\leq 2{\cal K} \\ s_{\bf q}}}  |B_{s_{\bf q}}|  \sum\limits_{\substack{{\bf k}, k\leq {\cal K}, |{\bf k} + {\bf q}|\leq {\cal K} \\ s_{\bf p}}}   | {\bf k}+{\bf q} | |b_{s_{\bf p}}({\bf k}+{\bf q})|    \sum\limits_{\substack{{\bf m}+{\bf l}+{\bf k}=0 \\ s_{\bf m}; s_{\bf l}; s_{\bf k} }} |b_{s_{\bf l}}| |B_{s_{\bf m}}|   | {C}_{{\bf k} {\bf l} {\bf m}}^{s_{\bf k} s_{\bf l} s_{\bf m}} |  \right] \mathrm{d}t  \\
\nonumber & \lesssim  & \frac{{\cal K}^2 \sqrt{\ln(2 {\cal K}L)} }{L^2 H^2  T}  \int_{t=0}^{t=T} \left[ \frac{ \| \omega \|_2 \|\bnabla {\bf v} \|_2}{\|{\bf v}\|_2^2}  \left(\| {\bf V} \|_2 \|\bnabla {\bf v} \|_2 +\| \omega \|_2 \| {\bf v} \|_2  \right)  \right] \mathrm{d}t \\
\nonumber & \lesssim  & \frac{{\cal K}^2 \sqrt{\ln(2 {\cal K}L)} }{L^2 H^2  T}  \int_{t=0}^{t=T} \left[ \frac{ \| \omega \|_2 \|\bnabla {\bf v} \|_2}{\|{\bf v}\|_2^2}  \left(  \| {\bf V} \|_2 \|\bnabla {\bf v} \|_2 +H \| \omega \|_2 \| \bnabla {\bf v} \|_2  \right)  \right] \mathrm{d}t \\
 \nonumber & \lesssim  & \frac{{\cal K}^2 \sqrt{\ln(2 {\cal K}L)} }{L^2 H^2  } \sup_t\left( \| {\bf V} \|_2 \| \omega \|_2 + H \| \omega \|_2^2 \right) \frac{1}{T} \int_{t=0}^{t=T}  \frac{ \|\bnabla {\bf v} \|_2^2 }{\|{\bf v}\|_2^2}  \mathrm{d}t \\
  & \lesssim  & \frac{{\cal K}^2 \ell  L^2 F^2 \sqrt{\ln(2 {\cal K}L)} }{ H  \nu^2} \left( 1+ \frac{H}{\ell}\right) \frac{1}{T} \int_{t=0}^{t=T}  \frac{ \|\bnabla {\bf v} \|_2^2 }{\|{\bf v}\|_2^2}  \mathrm{d}t \, ,\label{boundI2T2}
\end{eqnarray}
where we used the bounds on the suprema of the energy and enstrophy of the 2D base flow computed in appendix \ref{AppendixA}.

The first time integral in (\ref{ineqT2temp}) is bounded using the Cauchy-Schwarz inequality and (\ref{ineqtemp3}),
\begin{eqnarray}
\nonumber & & \frac{1}{T } \int_{t=0}^{t=T} \left[\frac{\nu {\cal K}^2}{  \| {\bf v} \|_2^2   } \sum\limits_{\substack{{\bf q}, q\leq 2{\cal K} \\ s_{\bf q}}}  |B_{s_{\bf q}}|  \sum\limits_{\substack{{\bf k}, k\leq {\cal K}, |{\bf k} + {\bf q}|\leq {\cal K} \\ s_{\bf p};s_{\bf k}}} |{\bf k}+{\bf q}| |b_{s_{\bf p}}({\bf k}+{\bf q})| k |b_{s_{\bf k}}| \right] \mathrm{d}t \\
& \lesssim & \nonumber  \frac{1}{T } \int_{t=0}^{t=T} \left[\frac{\nu {\cal K}^2}{  \| {\bf v} \|_2^2   } \sum\limits_{\substack{{\bf q}, q\leq 2{\cal K} \\ s_{\bf q}}}  |B_{s_{\bf q}}|  \frac{\| \bnabla {\bf v} \|_2^2}{L^2 H} \right] \mathrm{d}t \\
& \lesssim & \nonumber  \frac{1}{T } \int_{t=0}^{t=T} \left[\frac{\nu {\cal K}^2 \sqrt{\ln(2 {\cal K}L)}}{ L^2 H^{3/2}  } {\| \omega \|_2}    \frac{\| \bnabla {\bf v} \|_2^2}{ \| {\bf v} \|_2^2} \right] \mathrm{d}t \\
& \lesssim &  \nonumber \frac{\nu {\cal K}^2 \sqrt{\ln(2 {\cal K}L)}}{ L^2 H^{3/2}  } \sup_t\left(\| \omega \|_2 \right) \frac{1}{T } \int_{t=0}^{t=T}    \frac{\| \bnabla {\bf v} \|_2^2}{ \| {\bf v} \|_2^2}  \mathrm{d}t \\
& \lesssim &   \frac{ {\cal K}^2 F \sqrt{\ln(2 {\cal K}L)}}{ H  }  \frac{1}{T } \int_{t=0}^{t=T}    \frac{\| \bnabla {\bf v} \|_2^2}{ \| {\bf v} \|_2^2}  \mathrm{d}t \, ,\label{boundI1T2}
\end{eqnarray}
where we used once again the bound (\ref{boundSupomega}) on the supremum of the enstrophy.
Summing (\ref{boundI2T2}) and (\ref{boundI1T2}) gives the following bound on ${\cal T}_2$,
\begin{eqnarray}
\nonumber |{\cal T}_2| & \lesssim &  \frac{ {\cal K}^2 F \sqrt{\ln(2 {\cal K}L)}}{ H   } \left[ 1+ \left(1+ \frac{H}{\ell} \right)\left( \frac{L}{\ell} \right)^2 Gr \right]   \frac{1}{T } \int_{t=0}^{t=T}    \frac{\| \bnabla {\bf v} \|_2^2}{ \| {\bf v} \|_2^2}  \mathrm{d}t \\
& \lesssim & Gr \, \frac{(H+\ell) L^2 {\cal K}^2 F  \sqrt{\ln(2 {\cal K}L)}}{ \ell^3  H }   \frac{1}{T } \int_{t=0}^{t=T}    \frac{\| \bnabla {\bf v} \|_2^2}{ \| {\bf v} \|_2^2}  \mathrm{d}t \, .
\end{eqnarray}

From the bound (\ref{boundK}) on ${\cal K}$ and the bounds on the time-averaged enstrophy dissipation rate, one can easily see that ${\cal K} L$ has at most a power-law dependence in $Gr$ and $L/\ell$, and because we restrict attention to $Gr \geq 2$ we have
\begin{eqnarray}
\ln(2 {\cal K}L) \lesssim \ln Gr + \ln \frac {L}{\ell} \lesssim \ln \left(Gr \frac{L}{\ell} \right)\, .\label{boundlnKL}
\end{eqnarray}
Using this inequality and (\ref{boundK}) we finally bound $|{\cal T}_2|$ as
\begin{eqnarray}
|{\cal T}_2| \lesssim Gr^2 \frac{(H+\ell)  L^2 \la \| \bnabla \omega \|_2^2 \ra}{\ell^6    }   \ln^{3/2} \left(Gr \frac{L}{\ell} \right)   \frac{1}{T } \int_{t=0}^{t=T}    \frac{\| \bnabla {\bf v} \|_2^2}{ \| {\bf v} \|_2^2}  \mathrm{d}t \, .\label{boundT2}
\end{eqnarray}

We now wish to bound ${\cal T}_3$. From equation (\ref{eqv2}) and using H\"older's inequality,
\begin{eqnarray}
\left| \mathrm{d}_t \left(\|\textbf{v}\|_2^2 \right) \right| \leq 2 \| \bnabla {\bf V} \|_\infty  \| {\bf v} \|_2^2   + 2\nu \|\bnabla \textbf{v}\|_2^2 \, .
\end{eqnarray}
Using successively this inequality, (\ref{ineqkpg}), the Cauchy-Schwarz, (\ref{ineqtemp3}), and H\"older's inequality, we obtain
\begin{eqnarray}
| {\cal T}_3 | & \lesssim &  \frac{1}{T } \int_{t=0}^{t=T} \left[ \sum\limits_{\substack{{\bf p}+{\bf q}+{\bf k}=0 \\ k \leq {\cal K}; p \leq {\cal K}\\  s_{\bf p}; s_{\bf q}; s_{\bf k} }} | B_{s_{\bf q}}|   |b_{s_{\bf p}}|  |b_{s_{\bf k}}|    { k p} \, |{g}_{ {\bf q} {\bf k} {\bf p}}^{ s_{\bf q} s_{\bf k} s_{\bf p}}|  \frac{ \| \bnabla {\bf V} \|_\infty  \| {\bf v} \|_2^2   + \nu \|\bnabla \textbf{v}\|_2^2}{\| {\bf v} \|_2^4} \right] \mathrm{d}t \\
\nonumber & \lesssim &  \frac{{\cal K}^2}{ T } \int_{t=0}^{t=T} \left[ \frac{ \| \bnabla {\bf V} \|_\infty  \| {\bf v} \|_2^2   + \nu \|\bnabla \textbf{v}\|_2^2}{\| {\bf v} \|_2^4}  \sum\limits_{\substack{{\bf p}+{\bf q}+{\bf k}=0 \\ k \leq {\cal K}; p \leq {\cal K}\\ s_{\bf p}; s_{\bf q}; s_{\bf k} }}  | B_{s_{\bf q}}|   |b_{s_{\bf p}}|  |b_{s_{\bf k}}|   \right] \mathrm{d}t \\
\nonumber & \lesssim &  \frac{{\cal K}^2}{ L^2 H  T } \int_{t=0}^{t=T} \left[ \left( \| \bnabla {\bf V} \|_\infty + \nu \frac{ \|\bnabla \textbf{v}\|_2^2}{\| {\bf v} \|_2^2} \right) \sum\limits_{\substack{{\bf q}, q\leq 2{\cal K} \\ s_{\bf q}}}  |B_{s_{\bf q}}|    \right] \mathrm{d}t \\
\nonumber & \lesssim &  \frac{{\cal K}^2 \sqrt{\ln(2{\cal K} L)}}{ L^2 H^{3/2}  T } \int_{t=0}^{t=T} \| \omega \|_2  \left( \| \bnabla {\bf V} \|_\infty + \nu \frac{ \|\bnabla \textbf{v}\|_2^2}{\| {\bf v} \|_2^2} \right) \mathrm{d}t \\
\nonumber & \lesssim &  \frac{\nu {\cal K}^2 \sqrt{\ln(2{\cal K} L)}    }{  L^2 H^{3/2} }    \sup_t\left( \| \omega \|_2 \right)  \frac{1}{T} \int_{t=0}^{t=T}  \frac{ \|\bnabla \textbf{v}\|_2^2}{\| {\bf v} \|_2^2} \mathrm{d}t + \frac{{\cal K}^2\sqrt{\ln(2{\cal K} L)}}{ L^2 H^{3/2}  T}  \int_{t=0}^{t=T}  \| \omega \|_2 \| \bnabla {\bf V} \|_\infty  \mathrm{d}t \, .
\end{eqnarray}

Inserting the bound (\ref{boundSupomega}) on $\sup_t\left( \| \omega \|_2 \right)$, and the bounds (\ref{boundK}) and (\ref{boundlnKL}) on ${\cal K}$ and $\ln ( 2 {\cal K} L)$, we get
\begin{eqnarray}
\nonumber | {\cal T}_3 | & \lesssim &  \frac{  \la \| \bnabla \omega \|_2^2 \ra}{\ell^3  } Gr \ln^{3/2} \left( Gr  \frac{L}{\ell} \right) \frac{1}{T} \int_{t=0}^{t=T}  \frac{ \|\bnabla \textbf{v}\|_2^2}{\| {\bf v} \|_2^2} \mathrm{d}t \\
 & & +  \frac{ \la \| \bnabla \omega \|_2^2 \ra}{ L^2  \sqrt{H} \nu^2} \ln^{3/2} \left( Gr  \frac{L}{\ell} \right) \frac{1}{T}  \int_{t=0}^{t=T}  \| \omega \|_2 \| \bnabla {\bf V} \|_\infty  \mathrm{d}t \, .\label{boundT3}
\end{eqnarray}
We finally bound the integrated term ${\cal T}_4$: from (\ref{ineqkpg}), the Cauchy-Schwarz inequality and (\ref{ineqtemp3}), we have
\begin{eqnarray}
| {\cal T}_4 |&\lesssim &  \left| \frac{{\cal K}^2 }{L^2 H  T} \sup_t\left( \sum\limits_{\substack{{\bf q}, q\leq 2{\cal K} \\ s_{\bf q}}}  |B_{s_{\bf q}}|  \right)  \right| \\
&\lesssim & \frac{{\cal K}^2  \sqrt{\ln(2{\cal K} L)}  }{L^2 H^{3/2}  T} \sup_t \| \omega \|_2 \, ,
\end{eqnarray}
so this integrated term vanishes in the limit $T \to \infty$.

\section{Sufficient conditions for two-dimensionalization\label{suffcond}}

\subsection{Combining the bounds on the four terms}

Inserting (\ref{fourterms}) into (\ref{integrateineq}), with $K \geq \pi/H \geq 1/H$, leads to the following inequality,
\begin{eqnarray}
\nonumber \frac{1}{T} \ln  \frac{  \|\textbf{v}\|_2(t=T)}{ \|\textbf{v}\|_2(t=0)} & \leq & \frac{L^2 H^2}{2 \Omega} ( |{\cal T}_1|+|{\cal T}_2|+|{\cal T}_3|+|{\cal T}_4| ) \\
\nonumber & & - \frac{\nu}{4T} \int_{t=0}^{t=T} \frac{ \|\bnabla \textbf{v}\|_2^2}{\|\textbf{v}\|_2^2}\mathrm{d}t   +    \frac{1}{T} \int_{t=0}^{t=T} \lambda_1(t) \mathrm{d}t \, . 
\end{eqnarray}
Our goal is to show that the right-hand side of this inequality is negative in the limit $T \to \infty$, provided that $\Omega$ is large enough.
We substitute the bounds (\ref{boundT2}) and (\ref{boundT3}) on $|{\cal T}_2|$ and $|{\cal T}_3|$, which indicate that there are two positive dimensionless constants $c_2$ and $c_3$ such that
\begin{eqnarray}
 & & \frac{1}{T} \ln \frac{  \|\textbf{v}\|_2(t=T)}{ \|\textbf{v}\|_2(t=0)}  \leq \frac{1}{T} \int_{t=0}^{t=T} \lambda_1(t) \mathrm{d}t+ \frac{L^2 H^2}{2 \Omega}( |{\cal T}_1|+|{\cal T}_4| )  \label{replace2and3}\\
\nonumber &  & +  c_3  \frac{H^{3/2} \la \| \bnabla \omega \|_2^2 \ra}{ \Omega \nu^2} \ln^{3/2} \left( Gr  \frac{L}{\ell} \right) \frac{1}{T}  \int_{t=0}^{t=T}  \| \omega \|_2 \| \bnabla {\bf V} \|_\infty  \mathrm{d}t          \\
\nonumber & & + \left[ \frac{L^2 H^2 \la \| \bnabla \omega \|_2^2 \ra}{\ell^3 \Omega } Gr \ln^{3/2} \left( Gr  \frac{L}{\ell} \right) \left(c_3+c_2 Gr \frac{L^2 (H+\ell)}{\ell^3} \right)                 - \frac{\nu}{4} \right] \frac{1}{T}\int_{t=0}^{t=T} \frac{ \|\bnabla \textbf{v}\|_2^2}{\|\textbf{v}\|_2^2}\mathrm{d}t \, . 
\end{eqnarray}
Because we focus on $Gr \geq 2$, we have $c_3+c_2 Gr \frac{L^2 (H+\ell)}{\ell^3} \leq c_4 Gr \frac{L^2 (H+\ell)}{\ell^3}$, with $c_4=\max(c_2;c_3)$. Hence a sufficient condition for the square bracket in (\ref{replace2and3}) to be negative is
\begin{eqnarray}
\Omega \geq \Omega_1 = 4 c_4 \frac{L^4 H^2 (H+\ell)}{\ell^6} \frac{ \la \| \bnabla \omega \|_2^2 \ra}{\nu } Gr^2 \ln^{3/2} \left( Gr  \frac{L}{\ell} \right)  \, .\label{condition1}
\end{eqnarray}
When this condition is satisfied, one can drop the last term in (\ref{replace2and3}). Then upon taking the limit $T \to \infty$, using the Cauchy-Schwarz inequality for the middle line together with $\la \lambda_1(t) \ra = - \nu \pi^2/ (8H^2)$, $\lim_{T \to \infty}|{\cal T}_4|=0$, the bound (\ref{limT1}) on $\lim_{T \to \infty}|{\cal T}_1|$, and (\ref{boundGradVinf2}),
\begin{eqnarray}
& & \nonumber  \lim_{T \to \infty}  \frac{1}{T} \ln \frac{  \|\textbf{v}\|_2(t=T)}{ \|\textbf{v}\|_2(t=0)}   \leq  \la \lambda_1(t) \ra+  \frac{L^2 H^2}{2 \Omega} \left( \lim_{T \to \infty}|{\cal T}_1| +\lim_{T \to \infty}|{\cal T}_4| \right)\\
\nonumber &  &+    c_3  \frac{H^{3/2} \la \| \bnabla \omega \|_2^2 \ra}{ \Omega \nu^2} \ln^{3/2} \left( Gr  \frac{L}{\ell} \right) \sqrt{ \la \| \omega \|_2^2 \ra \la \| \bnabla {\bf V} \|_\infty^2 \ra }  \\
   & \leq & -\frac{\nu \pi^2}{8 H^2} +   \frac{c_1 H^2}{\nu^2   \Omega} \la \| \bnabla \omega \|_2^2 \ra \ln \left( Gr  \frac{L}{\ell} \right) \\
\nonumber & & \times  \left[ \frac{F L^2}{\ell^2} +  \left(\frac{1}{H} \sqrt{\la \|  \omega \|_2^2 \ra \la \| \bnabla \omega \|_2^2 \ra} + \la \| \bnabla \omega \|_2^2 \ra \right) \ln \left( Gr  \frac{L}{\ell} \right) \right]\, \\
\nonumber &  &+    c_3  \frac{H^{3/2} \la \| \bnabla \omega \|_2^2 \ra}{ \Omega \nu^2} \ln^{3/2} \left( Gr  \frac{L}{\ell} \right) \sqrt{ \la \| \omega \|_2^2 \ra \la \| \bnabla {\bf V} \|_\infty^2 \ra }  \\
& \leq & -\frac{\nu \pi^2}{8 H^2} +   \frac{ H^2}{\nu^2  \Omega} \la \| \bnabla \omega \|_2^2 \ra  \ln \left( Gr  \frac{L}{\ell} \right) \\
\nonumber & & \times  \left[ c_1 \frac{F L^2}{\ell^2} +  \left(\frac{c_1+c_3\sqrt{\tilde{c}_2}}{H} \sqrt{\la \|  \omega \|_2^2 \ra \la \| \bnabla \omega \|_2^2 \ra} + c_1 \la \| \bnabla \omega \|_2^2 \ra \right) \ln \left( Gr  \frac{L}{\ell} \right) \right]\, ,
\end{eqnarray}
where $c_1$ is a dimensionless constant. 
A sufficient condition for the right-hand side of this inequality to be negative is
\begin{eqnarray}
\nonumber \Omega \geq \Omega_2 & = & c_5 \frac{H^4  \la \| \bnabla \omega \|_2^2 \ra}{\nu^3}    \ln \left( Gr  \frac{L}{\ell} \right) \\
\nonumber & & \times \left[ \frac{F L^2}{\ell^2} +  \left(\frac{1}{H} \sqrt{\la \|  \omega \|_2^2 \ra \la \| \bnabla \omega \|_2^2 \ra} + \la \| \bnabla \omega \|_2^2 \ra \right) \ln \left( Gr  \frac{L}{\ell} \right) \right] \, , \\
\label{condition2}
\end{eqnarray}
where $c_5=c_1+c_3\sqrt{\tilde{c}_2}$. \\

To summarize, if $\Omega> \Omega_1$ and $\Omega> \Omega_2$, where $\Omega_1$ and $\Omega_2$ are defined in (\ref{condition1}) and (\ref{condition2}) respectively, then
\begin{eqnarray}
\lim_{T \to \infty}  \frac{1}{T} \ln \frac{  \|\textbf{v}\|_2(t=T)}{ \|\textbf{v}\|_2(t=0)}  < 0 \, ,
\end{eqnarray}
and therefore $\lim_{t \to \infty} \|\textbf{v}\|_2(t) = 0$: the 2D base flow is linearly stable to 3D perturbations.

As a matter of fact, using the bounds (\ref{boundEnstrophyF}) and (\ref{boundEDF}) one can prove that $\Omega_2 \lesssim \Omega_1 \ln^{1/2} \left( Gr \frac{L}{\ell} \right)$. Hence $\Omega \geq c \Omega_1 \ln^{1/2} \left( Gr \frac{L}{\ell} \right)$ is a sufficient condition to have both $\Omega \geq \Omega_1$ and $\Omega \geq \Omega_2$.
We therefore obtain the following sufficient condition for the flow to be linearly stable to 3D perturbations,
\begin{eqnarray}
\Omega \geq \Omega_3 = c_6  \frac{ \la \| \bnabla \omega \|_2^2 \ra}{\nu } Gr^2 \ln^{2} \left( Gr  \frac{L}{\ell} \right) \frac{L^4 H^2 (H+\ell) }{\ell^6} \, , \label{condition3}
\end{eqnarray}
where $c_6$ is a dimensionless constant.

\subsection{Criteria based on the root-mean-square velocity}
The sufficient condition (\ref{condition3}) indicates unambiguously that the flow is linearly stable to three-dimensional perturbations above a critical value of the rotation rate $\Omega$. However, the quantity $\la \| \bnabla \omega \|_2^2 \ra$ appearing in this criterion is difficult to measure or evaluate. We therefore use the bounds (\ref{boundED}) and (\ref{boundEDKolmogorov}) on the time-averaged enstrophy dissipation rate, together with (\ref{boundGr}) and (\ref{boundln}), to produce alternate sufficient conditions for two-dimensionalization that are expressed in terms of the r.m.s. velocity $U$. 

The bound (\ref{boundED}) gives
\begin{eqnarray}
 \frac{\ell \Omega_3}{U} \lesssim Re^6 \ln^{2} \left( Re  \frac{L}{\ell} \right)  \frac{L^6 H^3 (H+\ell)}{\ell^{10}}    \, ,
\end{eqnarray}
hence the two-dimensional flow is stable if the inverse Rossby number is greater than the right-hand side of this inequality, i.e., if
\begin{eqnarray}
 Ro \leq Ro_< = c_7 Re^{-6} \ln^{-2} \left( Re  \frac{L}{\ell} \right)  \frac{\ell^{10}}{L^6 H^3 (H+\ell)}     \, , \label{conditionRo}
\end{eqnarray}
where $c_7$ is a dimensionless constant that depends on the shape of the forcing only.

A somewhat less stringent criterion for two-dimensionalization can be obtained when the forcing is of single-mode type: bounding $\la \| \bnabla \omega \|_2^2 \ra$ in $\Omega_3$ using (\ref{boundEDKolmogorov}), we obtain the following sufficient condition for the decay of infinitesimal 3D perturbations,
\begin{eqnarray}
Ro \leq Ro_<^{(\text{SM})} = c_8  Re^{-5} \ln^{-2} \left( Re  \frac{L}{\ell} \right)  \frac{\ell^{10}}{L^6 H^3 (H+\ell)}  \, ,\label{conditionRoKolmogorov}
\end{eqnarray}
where $c_8$ is yet another dimensionless constant that depends on the shape of the forcing only.

\subsection{Criterion based on the forcing strength}

We can derive an alternate sufficient criterion for two-dimensionalization in terms of the forcing strength instead of the r.m.s. velocity. Inserting the upper bound (\ref{boundEDF}) on the time-averaged enstrophy dissipation rate into the expression for $\Omega_3$ leads to a  criterion in terms of the Grashof and forcing Rossby numbers:
\begin{eqnarray}
 Ro^{(f)} \leq Ro_<^{(f)} = c_9 Gr^{-7/2} \ln^{-2} \left( Gr  \frac{L}{\ell} \right)  \frac{\ell^{10}}{L^6 H^3 (H+\ell)}     \, , \label{conditionRoF}
\end{eqnarray}
where $c_9$ is a dimensionless constant that depends on the shape of the forcing only. When (\ref{conditionRoF}) is satisfied, the global attractor of the 2D Navier-Stokes equation is stable to infinitesimal 3D perturbations.

\section{Absolute two-dimensionalization \label{abstwodim}}

We now prove two-dimensionalization starting from arbitrary initial conditions: the initial 2D part of the flow is not necessarily on the attractor of the 2D Navier-Stokes equation, and the vertically-dependent part of the velocity field can be arbitrarily large initially. We use a theorem proven by Babin, Mahalov and Nicolaenko on the existence of a global attractor for the 3D rotating Navier-Stokes equations, when the rotation rate is sufficiently large \citep{Babin2}. For the setup considered here, their theorem 1 says the following: consider given values of $F$, $\nu$, $\ell$, $L$, $H$, a smooth shape of the forcing function and some $\alpha>1/2$. Then above some value $\Omega'(F,\nu,\ell,L,H,\alpha)$ of the rotation rate, and for any initial condition ${\bf u}_0={\bf u}(t=0)$, there is a time $T_0$ after which spatial derivatives of ${\bf u}$ of order up to $\alpha$ are bounded in the $L_2$ sense by a constant that is independent of time, of $\Omega$, and of the initial condition ${\bf u}_0$. The time  $T_0$ does depend on ${\bf u}_0$, $F$, $\nu$, $\ell$, $L$ and $H$. We stress the fact that $\Omega'$ is independent of the initial condition ${\bf u}_0$. For large times, $t>T_0$, and provided that $\Omega > \Omega'$, we have
\begin{eqnarray}
\sup_{t>T_0} \| \bnabla^\beta {\bf u}  \|_2 < C(\alpha, F, \nu, \ell, L, H) \, , \label{thBMN}
\end{eqnarray}
for  $0 \leq \beta \leq \alpha$.
The key point is that the quantity $C$ on the right-hand side is independent of $\Omega$ and of the initial condition ${\bf u}_0$. It depends only the parameters indicated above, as well as on the shape function of the forcing.  In the following, we only need control over the first few derivatives of ${\bf u}$, so we now fix $\alpha$ to some large value (for instance $\alpha=10$).

We wish to prove that, for any value of the Grashof number, there is a threshold value of the forcing Rossby number below which the 3D part of the velocity field decays at least exponentially in the long-time limit, regardless of its initial amplitude: the flow two-dimensionalizes in the long-time limit, starting from arbitrarily large 3D perturbations. We call this property absolute two-dimensionalization.

We only wish to prove the existence of the threshold forcing Rossby number for absolute two-dimensionalization, and we do not determine its precise dependence on the Grashof number and geometry of the system. Therefore, to alleviate the algebra, in the following we denote as $C$ any quantity that is independent of time, of $\Omega$, and of the initial condition ${\bf u}_0$. A quantity denoted as $C$ is only a function of $F$, $\nu$, $\ell$, $L$, and $H$. $C$ denotes the generic such quantity: it is not necessarily dimensionless, and it can change value and dimension from one line of algebra to the next. By contrast, numbered quantities $C_i$ are similar to $C$ but keep the same value throughout the algebra.

We are interested in the long-time behavior of the system, when global rotation is fast. In the following, we therefore restrict attention to $\Omega>\Omega'$, and to $t>T_0$. Under these conditions, we can use the inequality (\ref{thBMN}) above to bound the $L_2$ norm of velocity gradients of order up to $\alpha$ by $\Omega$-independent quantities. Using Agmon's inequality, (\ref{thBMN}) also implies that the $L_\infty$ norm of gradients of ${\bf u}$ is bounded by some $\Omega$-independent and time-independent quantity $C$:

\begin{eqnarray}
\sup_{t>T_0} \| \bnabla^\beta {\bf u}  \|_\infty < C(\alpha, F, \nu, \ell, L, H) \, , \label{thBMNinfty}
\end{eqnarray}
for $0 \leq \beta \leq \alpha-2$. Because we have fixed $\alpha$ to some large value, in the following we bound the spatial derivatives of ${\bf u}$ using (7.2) when needed.

\subsection{Evolution of the three-dimensional part of the velocity field\label{StratAbs}}

We consider the evolution of an initially 3D velocity field under the fully nonlinear 3D rotating Navier-Stokes equation (\ref{NS}). Define the vertical average of a field ${\bf h}$ as
\begin{equation}
\bar{{\bf h}}(x,y,t) = \frac{1}{H} \int_{z=0}^{z=H}  {\bf h}(x,y,z,t) \, \mathrm{d}z \, ,
\end{equation}
and decompose the velocity field into 
\begin{equation}
\textbf{u}=\textbf{V}(x,y,t)+\textbf{v}(x,y,z,t) \mbox{, with } \textbf{V}=\bar{\textbf{u}} \mbox{, and } \textbf{v}=\textbf{u}-\textbf{V} \, .
\label{decomp}
\end{equation}
In the following we call ${\bf V}$ (resp. ${\bf v}$) the two-dimensional (resp. three-dimensional) part of the velocity field. In contrast with the previous sections, here the 3D part ${\bf v}$ of the velocity field need not be small. Our goal is to find a criterion that insures the decay of ${\bf v}$. The evolution equation for the two-dimensional part of the velocity field is obtained by vertically-averaging  equation (\ref{NS}),
\begin{equation}
\partial_t \textbf{V} +(\textbf{V} \cdot \bnabla) \textbf{V} +\overline{(\textbf{v} \cdot \bnabla) \textbf{v}}=-\bnabla{P} + \nu \Delta \textbf{V} + \textbf{f}(x,y) \, ,
\label{eqVabs}
\end{equation}
where $P$ includes the vertically averaged pressure and the vertically averaged Coriolis force, which is a gradient.

By subtracting this equation to the rotating Navier-Stokes equation, we get the evolution equation for the three-dimensional part of the flow:
\begin{equation}
\partial_t \textbf{v} +(\textbf{V} \cdot \bnabla) \textbf{v} +(\textbf{v} \cdot \bnabla) \textbf{V} +(\textbf{v} \cdot \bnabla) \textbf{v} - \overline{(\textbf{v} \cdot \bnabla) \textbf{v}} + 2 \Omega {\bf e}_z \times \textbf{v} =-\bnabla(p-P) + \nu \Delta \textbf{v}   \, .
\label{eqvabs}
\end{equation}
Incompressibility imposes $\bnabla \cdot \textbf{V} = 0$ and $\bnabla \cdot \textbf{v} = 0$.
 
 \subsection{Control over the small scales of the 3D part of the velocity field}
 
One can obtain the same equation as (\ref{eqlnv}) for the evolution of the kinetic energy of the 3D part of the velocity field, where ${\bf V}$ and ${\bf v}$ are now solutions of (\ref{eqVabs}) and (\ref{eqvabs}). Like in section \ref{largevssmall}, we introduce a cut-off wavenumber $\Kabs$ and decompose the 3D part of the velocity field into ${\bf v}={\bf v}_<+{\bf v}_>$, distinguishing between wave numbers that are lower or greater than $\Kabs$. We then repeat the steps (\ref{eqlnvdecomp}) to (\ref{ineqlnvLS}), and obtain 
\begin{eqnarray}
\nonumber \frac{1}{2} \mathrm{d}_t \left( \ln  \|\textbf{v}\|_2^2 \right) & \leq &   \frac{1}{\|\textbf{v}\|_2^2} \left[- \int_\mathcal{D} \textbf{v}_< \cdot (\bnabla \textbf{V}) \cdot \textbf{v}_<  \, \mathrm{d}^3 \textbf{x}  - \frac{\nu}{4} \|\bnabla \textbf{v}\|_2^2 \right] +   \frac{4}{\nu {\Kabs}^2} \| \bnabla {\bf V} \|_\infty^2 - \frac{\nu \pi^2}{4 H^2} \, . \label{lnvabs1}\\
\end{eqnarray}
We choose the following value of the cut-off wavenumber:
\begin{eqnarray}
\Kabs = \frac{4 H}{\nu \pi} \sup_{t>T_0} \| \bnabla {\bf V} \|_\infty \, , \label{valKabs}
\end{eqnarray}
which we know is a finite quantity thanks to equation (\ref{thBMNinfty}).
This ensures that the sum of the last two terms in (\ref{lnvabs1}) is negative and can be discarded (again for $t>T_0$), which leads to
\begin{eqnarray}
 \frac{1}{2} \mathrm{d}_t \left( \ln  \|\textbf{v}\|_2^2 \right) & \leq &   \frac{1}{\|\textbf{v}\|_2^2} \left[- \int_\mathcal{D} \textbf{v}_< \cdot (\bnabla \textbf{V}) \cdot \textbf{v}_<  \, \mathrm{d}^3 \textbf{x}  - \frac{\nu}{4} \|\bnabla \textbf{v}\|_2^2 \right]  \, . \label{lnvabs2}   
\end{eqnarray}

From (\ref{thBMNinfty}), the cut-off wave number (\ref{valKabs}) is bounded by an $\Omega$-independent and time-independent quantity:
\begin{eqnarray}
\Kabs \leq  \frac{4 H}{\nu \pi} \sup_{t>T_0} \| \bnabla {\bf u} \|_\infty \leq C \, .\label{boundKabs}
\end{eqnarray}

\subsection{Control over the large scales of the 3D part of the velocity field}

Decompose ${\bf v}$ and ${\bf V}$ on the basis of helical modes as in (\ref{helical}) and (\ref{helicalV}). Inserting these decompositions into the evolution equation (\ref{eqvabs}) and projecting onto each helical mode leads to the evolution equation for the helical amplitudes of ${\bf v}$: 
\begin{eqnarray}
\nonumber \partial_t b_{s_{\bf k}} & = & -\nu k^2 b_{s_{\bf k}} + \sum\limits_{\substack{{\bf p}+{\bf q}+{\bf k}=0 \\ s_{\bf p}; s_{\bf q} }}  C_{{\bf k} {\bf p} {\bf q}}^{s_{\bf k} s_{\bf p} s_{\bf q}} \, \left[ b_{s_{\bf p}}^* B_{s_{\bf q}}^* e^{-i(\sigma_{s_{\bf p}}+\sigma_{s_{\bf k}})t}  + \frac{1}{2} b_{s_{\bf p}}^* b_{s_{\bf q}}^* e^{-i(\sigma_{s_{\bf p}}+\sigma_{s_{\bf k}}+\sigma_{s_{\bf q}})t} \right]\, , \\  \label{eqbskabs}  
\end{eqnarray}
where $C_{{\bf k} {\bf p} {\bf q}}^{s_{\bf k} s_{\bf p} s_{\bf q}}$ is still given by (\ref{defCkpq}).
Using the vorticity equation for the 2D part of the flow, we get the following evolution equation for the helical amplitudes of ${\bf V}$:
\begin{eqnarray}
\nonumber \partial_t B_{s_{\bf k}} & = & \frac{-s_{\bf k}}{\sqrt{2} k} \left[ -i {\bf k}\cdot \{ \omega {\bf V} \}_{\bf k} + (i {\bf k} \times \{ \overline{ {\bf v} \times (\bnabla \times {\bf v})    }  \}_{\bf k})\cdot {\bf e}_z -\nu k^2 \omega_{\bf k} + (i{\bf k} \times {\bf f}_{\bf k})\cdot {\bf e}_z  \right] \, , \\
\label{eqBskabs}
\end{eqnarray}
where ${\bf f}_{\bf k}$ and $\omega_{\bf k}$ are the Fourier amplitudes of ${\bf f}$ and $\omega$, using Fourier series similar to (\ref{Fourierv}) and (\ref{FourierV}). Similarly, $\{ \dots \}_{\bf k}$ denotes the Fourier amplitude of its argument on the wave vector ${\bf k}$.

In contrast with the proof of linear two-dimensionalization, here the 3D part of the velocity field contains many Fourier modes in $z$, the evolution of which is coupled. 
Writing the vertical component of ${\bf k}$ as $k_z= \frac{n \pi}{H}$ with $n \in \mathbb{Z}_{\neq 0}=\mathbb{Z} \setminus \{ 0 \}$, the triple velocity product becomes
\begin{eqnarray}
& & \int_\mathcal{D} \textbf{v}_< \cdot (\bnabla \textbf{V}) \cdot \textbf{v}_<  \, \mathrm{d}^3 \textbf{x} \label{tripleproductabs} \\
\nonumber & & = - L^2 H \sum\limits_{ n \in \mathbb{Z}_{\neq 0}  }  \sum\limits_{\substack{{\bf p}+{\bf q}+{\bf k}=0  \\ k_z= \frac{n \pi}{H}; k \leq {\Kabs}; p \leq {\Kabs}\\ s_{\bf p}; s_{\bf q}; s_{\bf k} }} B_{s_{\bf q}} b_{s_{\bf k}} b_{s_{\bf p}}  e^{i(\sigma_{s_{\bf k}}+\sigma_{s_{\bf p}})t} s_{\bf k} s_{\bf p} k p \frac{\sigma_{s_{\bf k}}+\sigma_{s_{\bf p}}}{2 \Omega k_z} g_{ {\bf q} {\bf k} {\bf p}}^{*  s_{\bf q} s_{\bf k} s_{\bf p}} \\
\nonumber & & =  - \frac{L^2 H^2}{\pi} \sum\limits_{ n \in \mathbb{Z}_{\neq 0}  }  \frac{1}{n} \sum\limits_{\substack{{\bf p}+{\bf q}+{\bf k}=0  \\ k_z= \frac{n \pi}{H}; k \leq {\Kabs}; p \leq {\Kabs}\\ s_{\bf p}; s_{\bf q}; s_{\bf k} }} B_{s_{\bf q}} b_{s_{\bf k}} b_{s_{\bf p}}  e^{i(\sigma_{s_{\bf k}}+\sigma_{s_{\bf p}})t} s_{\bf k} s_{\bf p} k p \frac{\sigma_{s_{\bf k}}+\sigma_{s_{\bf p}}}{2 \Omega} g_{ {\bf q} {\bf k} {\bf p}}^{*  s_{\bf q} s_{\bf k} s_{\bf p}} \, ,
\end{eqnarray}
which differs from (\ref{tripleproduct}) by an infinite sum over the vertical Fourier modes of ${\bf v}$.

Integrate (\ref{lnvabs2}) from time $t=T_0$ to $t=T_0+T$ and divide by $T$ to obtain
\begin{eqnarray}
 \frac{1}{T} \ln  \frac{  \|\textbf{v}\|_2(t=T_0+T)}{ \|\textbf{v}\|_2(t=T_0)} & \leq & - \frac{1}{T }\int_{t=T_0}^{t=T_0+T} \frac{\int_\mathcal{D} \textbf{v}_< \cdot (\bnabla \textbf{V}) \cdot \textbf{v}_<  \, \mathrm{d}^3 \textbf{x}}{\| {\bf v} \|_2^2} \mathrm{d}t  \label{integrateineqabs}\\
\nonumber & &   - \frac{\nu}{4T} \int_{t=T_0}^{t=T_0+T} \frac{ \|\bnabla \textbf{v}\|_2^2}{\|\textbf{v}\|_2^2}\mathrm{d}t   \, . 
\end{eqnarray}

Insert the expression (\ref{tripleproductabs}) of the triple velocity product, before performing an integration by parts in time, integrating the oscillatory exponential and differentiating the rest of the integrand. This gives
\begin{eqnarray}
\nonumber  - \frac{1}{T }\int_{t=T_0}^{t=T_0+T} \frac{\int_\mathcal{D} \textbf{v}_< \cdot (\bnabla \textbf{V}) \cdot \textbf{v}_<  \, \mathrm{d}^3 \textbf{x}}{\| {\bf v} \|_2^2} \mathrm{d}t & = & \frac{L^2 H^2}{2 \pi \Omega} \sum\limits_{ n \in \mathbb{Z}_{\neq 0}  }  \frac{1}{n} \left( {\cal T}_1^{(n)} + {\cal T}_2^{(n)} +{\cal T}_3^{(n)} + {\cal T}_4^{(n)} \right)\, , \\
 \label{fourtermsabs}
\end{eqnarray}
where ${\cal T}_1^{(n)}$, ${\cal T}_2^{(n)}$, ${\cal T}_3^{(n)}$ and ${\cal T}_4^{(n)}$ are given by expressions identical to (\ref{defT1})-(\ref{defT4}), only for two minor changes: the bounds for the time integrations (and for the boundary terms of ${\cal T}_4^{(n)}$) are now $T_0$ to $T_0+T$, and the sums are now restricted to wave vectors such that $k_z=n \pi /H$, $k\leq \Kabs$ and $p\leq \Kabs$ (see for instance the second sum on the right-hand side of (\ref{tripleproductabs})).

Using inequalities (\ref{thBMN}) and (\ref{thBMNinfty}), we prove in appendix \ref{secBabs} that:
\begin{eqnarray}
 \left| \sum\limits_{ n \in \mathbb{Z}_{\neq 0}  }  \frac{1}{n}  {\cal T}_1^{(n)} \right| & \leq & C_1 \, , \label{boundT1abs}\\
 \left| \sum\limits_{ n \in \mathbb{Z}_{\neq 0}  }  \frac{1}{n}  {\cal T}_2^{(n)} \right| & \leq & C_2 \, , \label{boundT2abs}\\
 \left| \sum\limits_{ n \in \mathbb{Z}_{\neq 0}  }  \frac{1}{n}  {\cal T}_3^{(n)} \right| & \leq & C_3 + \frac{C_4}{T}  \int_{t=T_0}^{t=T_0+T} \frac{ \|\bnabla \textbf{v}\|_2^2}{\|\textbf{v}\|_2^2}\mathrm{d}t \label{boundT3abs}\\
\lim_{T \to \infty} \left| \sum\limits_{ n \in \mathbb{Z}_{\neq 0}  }  \frac{1}{n}  {\cal T}_4^{(n)} \right| & = & 0 \label{boundT4abs} \, ,
\end{eqnarray}
where the $C_i$ are $\Omega$-independent and $T$-independent quantities. Inserting these bounds in (\ref{fourtermsabs}) we obtain
\begin{eqnarray}
\left| \frac{1}{T }\int_{t=T_0}^{t=T_0+T} \frac{\int_\mathcal{D} \textbf{v}_< \cdot (\bnabla \textbf{V}) \cdot \textbf{v}_<  \, \mathrm{d}^3 \textbf{x}}{\| {\bf v} \|_2^2} \mathrm{d}t \right| & \leq & \frac{C_5}{\Omega} + \frac{L^2 H^2}{2 \pi \Omega} \left| \sum\limits_{ n \in \mathbb{Z}_{\neq 0}  }  \frac{1}{n}  {\cal T}_4^{(n)} \right|  \label{tempabs}\\
\nonumber &  & +  \frac{L^2 H^2 C_4}{2 \pi \Omega T}  \int_{t=T_0}^{t=T_0+T} \frac{ \|\bnabla \textbf{v}\|_2^2}{\|\textbf{v}\|_2^2}\mathrm{d}t \, .
\end{eqnarray}
Using Poincar\'e's inequality on (half of) the viscous term in (\ref{integrateineqabs}), we obtain
\begin{eqnarray}
- \frac{\nu}{4T} \int_{t=T_0}^{t=T_0+T} \frac{ \|\bnabla \textbf{v}\|_2^2}{\|\textbf{v}\|_2^2}\mathrm{d}t \leq  - \frac{\nu \pi^2}{8H^2}    - \frac{\nu}{8T} \int_{t=T_0}^{t=T_0+T} \frac{ \|\bnabla \textbf{v}\|_2^2}{\|\textbf{v}\|_2^2}\mathrm{d}t \, , \label{boundviscabs}
\end{eqnarray}
and inserting (\ref{tempabs}) and (\ref{boundviscabs}) in (\ref{integrateineqabs}) gives
\begin{eqnarray}
\nonumber   \frac{1}{T} \ln  \frac{  \|\textbf{v}\|_2(t=T_0+T)}{ \|\textbf{v}\|_2(t=T_0)} & \leq & \left( \frac{C_5}{\Omega}  - \frac{\nu \pi^2}{8H^2} \right) + \left( \frac{L^2 H^2 C_4}{2 \pi \Omega} - \frac{\nu}{8} \right) \frac{1}{T}  \int_{t=T_0}^{t=T_0+T} \frac{ \|\bnabla \textbf{v}\|_2^2}{\|\textbf{v}\|_2^2}\mathrm{d}t  \\
  & & +  \frac{L^2 H^2}{2 \pi \Omega} \left| \sum\limits_{ n \in \mathbb{Z}_{\neq 0}  }  \frac{1}{n}  {\cal T}_4^{(n)} \right| \, . \label{finalabs}
\end{eqnarray}
If $\Omega$ is large enough to satisfy
\begin{eqnarray}
\Omega > C_6  = \max \left\{  \frac{8 C_5 H^2}{\nu \pi^2} ;  \frac{4 L^2 H^2 C_4}{ \pi \nu}  \right\} \, ,\label{Oabstemp}
\end{eqnarray}
then the two parentheses on the right-hand side of (\ref{finalabs}) are negative, and these two first terms can be discarded. Upon taking the limit $T\to \infty$ and using (\ref{boundT4abs}), we finally obtain
\begin{eqnarray}
\lim_{T \to \infty}  \frac{1}{T} \ln  \frac{  \|\textbf{v}\|_2(t=T_0+T)}{ \|\textbf{v}\|_2(t=T_0)}  <  0 \, ,
\end{eqnarray}
which shows that the three-dimensional part of the velocity field decays in the long-time limit.

To summarize, we have proven that if $\Omega$ is larger than the value $C_6 (F,\nu,\ell,H,L)$ defined in (\ref{Oabstemp}), then the flow becomes 2D in the long-time limit, regardless of the initial condition, and even if the 3D part of the velocity field is arbitrarily large initially. Recall that we also restricted attention to $\Omega \geq \Omega'$, to use (\ref{thBMN}) and (\ref{thBMNinfty}). In terms of dimensionless parameters, this proves the existence of a finite threshold value $Ro^{(f)}_{\text{abs}} (Gr,\ell/L, H/L)$ of the forcing Rossby number under which {\it absolute two-dimensionalization} takes place: for $Ro^{(f)} < Ro^{(f)}_{\text{abs}} (Gr,\ell/L, H/L)$, the flow two-dimensionalizes in the long-time limit, regardless of the initial condition ${\bf u}_0$ for the velocity field.

\section{Discussion\label{discsection}}

The central result of the present study is the fact that rapidly rotating flows driven by a vertically invariant horizontal body force admit  exactly two-dimensional solutions that are robust to 3D perturbations. We have first considered the 3D rotating Navier-Stokes equation linearized around a 2D and possibly turbulent base flow, to prove linear two-dimensionalization: defining a Reynolds number $Re$ and a Rossby number $Ro$ based on the r.m.s. velocity, we have determined a region of the parameter space $(Re,Ro)$ where the (possibly turbulent) two-dimensional flow is linearly stable to three-dimensional perturbations. More precisely, for any Reynolds number, there is a threshold value of the Rossby number $Ro_c(Re)$ under which the flow ends up in the attractor of the 2D Navier-Stokes equation, provided that the initial 3D perturbations are weak. Similar results have been obtained in terms of the control parameters of the system: there is a Grashof-number-dependent threshold value of the forcing-based Rossby number $Ro^{(f)}_c(Gr)$ under which the flow ends up in the attractor of the 2D Navier-Stokes equation, provided that the initial 3D perturbations are weak.

We then came back to the fully nonlinear 3D rotating Navier-Stokes equation and proved absolute two-dimensionalization: under a Grashof-number-dependent threshold value $Ro^{(f)}_\text{abs}(Gr)$ of the forcing-based Rossby number, the flow becomes 2D in the long-time limit, regardless of the initial condition, including initial velocity fields that have very strong vertically dependent structures.

To summarize, when $Ro^{(f)}< Ro^{(f)}_c$, our study indicates that the global attractor of the 2D Navier-Stokes equation is an attractor of the 3D rotating system. If $Ro^{(f)}_\text{abs} \leq Ro^{(f)} \leq Ro^{(f)}_c$,  this attractor may coexist in phase space with fully 3D attractors, and the fate of the flow may depend on the initial condition. By contrast, when rotation is so fast that $Ro^{(f)}< Ro^{(f)}_\text{abs}$, the global attractor of the 2D Navier-Stokes equation is the only attractor of the 3D rotating system, and any initial condition becomes exactly 2D in the long-time limit.

We have determined conservative lower bounds on $Ro_c(Re)$, given by $Ro_<$ in (\ref{conditionRo}) for generic time-independent forcing, and by $Ro_<^{(\text{SM})}$ in (\ref{conditionRoKolmogorov}) for single-mode forcing. Similarly, we obtained a lower bound on $Ro^{(f)}_c(Gr)$, given by $Ro_<^{(f)}(Gr)$ in (\ref{conditionRoF}). These lower bounds might be improved using finer estimates and inequalities. The rapid decrease of the lower bounds $Ro_<$ and $Ro_<^{(\text{SM})}$ (resp. $Ro_<^{(f)}(Gr)$) with increasing Reynolds (resp. Grashof) number stems from the fact that small-scale perturbations are more difficult to control when viscosity is low. How fast the true threshold-Rossby-numbers $Ro_c(Re)$ really decreases with $Re$ could be determined through careful numerical simulations. These bounds also indicate unambiguously that two-dimensionalization is more easily achieved in shallow domains: as $H$ decreases with fixed $L$ and $\ell$, the bounds $Ro_<$, $Ro_<^{(\text{SM})}$ and $Ro_<^{(f)}$ increase as $H^{-3}$ and therefore the criterion for two-dimensionalization is more easily met (using another bounding strategy than the one presented here, the better scaling $H^{-5}$ can be obtained, but at the expense of the scalings in $Re$ or $Gr$). This result is qualitatively compatible with the numerical observations of \citet{Smith} and \citet{Deusebio}, who report that, for fixed rotation rate, the large scales of turbulent flows in shallow domains are approximately 2D  and exhibit an inverse energy cascade, whereas in deep domains the system displays no inverse cascade. The comparison with these studies is only qualitative, for two reasons: first, they are focused on the transient behavior only and they do not consider the statistically steady state attained in the long-time limit. Second, they consider the two-dimensionalization of the large scales of the flow only, whereas we are interested in exact two-dimensionalization, where the system becomes 2D at every horizontal scale.

The fact that the attractor of the 2D Navier-Stokes equation (\ref{NS2D}) is an attractor for the 3D rotating flow when rotation is fast enough allows to give precise answers to the questions raised at the outset of this study: 
\begin{itemize}
\item First, in the two-dimensional attractor, the system exhibits no energy dissipation anomaly. As proven by Alexakis and Doering (see also appendix \ref{AppendixA}), the time-averaged energy dissipation per unit mass $\epsilon$ is bounded from above by $\epsilon \lesssim Re^{-1/2} \frac{U^3}{\ell}$ and therefore $\epsilon \ell / U^3$ vanishes in the infinite Reynolds number limit. 
\item Second, these results shed some light on the decrease of intermittency with global rotation: experimental and numerical studies of the moments of the velocity structure functions report that intermittency is reduced by global rotation \citep{Baroud, Muller, Seiwert, Mininni}. For a given large value of the Reynolds number and without global rotation, the body-forced flows considered in this study are three-dimensional, and most likely turbulent and intermittent \citep{Frisch}. With global rotation, however, if the Rossby number is lower than $Ro_c(Re)$ the flow has a (possibly turbulent) 2D attractor, and the belief is that the energy transfers of such 2D flows do not display intermittency \citep{Paret, Boffetta}.
\item Third, the system has a symmetric vorticity distribution: cyclone-anticyclone asymmetry disappears completely in this 2D attractor. Indeed, in the 2D Navier-Stokes equations the Coriolis force is absorbed by the pressure gradient and $\Omega$ disappears from the equation: there is no preferred direction of rotation and no asymmetry of the vorticity distribution.
\end{itemize}

Finally, these results can be discussed in the context of wave turbulence. For the last decade, there has been a debate over whether rapidly rotating turbulence can be described in terms of weak turbulence of inertial waves \citep{Galtier, Cambon, Yarom, Scott}. This theory describes the energy transfers of rapidly rotating turbulence through the weakly nonlinear three-wave interaction of inertial waves. Such a weak nonlinearity assumption is satisfied if the period of the inertial waves is much shorter than the characteristic time of nonlinear energy transfers between waves. In this framework, purely two-dimensional modes -- the 2D ``vortex" mode -- are therefore discarded, because they correspond to vanishing inertial frequency (a recent attempt to include them is described in \citet{Scott}, but this study only considers 2D flows at much larger scales than the waves: this is the limit of an almost uniform 2D flow Doppler-shifting the waves). As mentioned in the introduction, a key point of the theory is that three-wave interactions do not transfer energy from the inertial waves to the 2D vortex mode: if the latter is zero initially, it remains zero through three-wave interactions. This is only valid for short times, however, because for longer times -- or equivalently, at next order in the weakly nonlinear expansion -- quasi-resonances and quartets of inertial waves come into play, and can transfer energy from the three-dimensional waves to two-dimensional motion \citep{SmithWaleffe}.

Weak inertial-wave turbulence therefore seems to be a good candidate to describe the initial evolution of a rapidly rotating 3D flow, provided the initial condition contains negligible energy in purely 2D  modes. However, it is unlikely to apply to stationary rotating turbulence. As a matter of fact, experimental and numerical studies of stationary rotating turbulence depart strongly from the predictions of wave turbulence, with a large fraction of the energy of the system accumulating in the two-dimensional vortex mode \citep{Baroud,Alexakis,Gallet2014,CampagnePOF, CampagnePRE}. An explanation that is often given is that the Rossby number of these studies is not low enough to enter the weakly nonlinear regime where inertial-wave turbulence would apply. However, we believe that energetic two-dimensional motions are a generic feature of stationary rotating turbulence in bounded domains, which challenges the applicability of weak-turbulence theory.

Indeed, one can think of two extreme types of body-forced turbulence: the case where the forcing inputs energy in 3D modes only, and the case where it inputs energy into 2D motion only. The first situation is addressed in the thorough numerical study of \citet{Alexakis}, who considers 3D forcing of the Taylor-Green type: in the rapidly rotating and low-viscosity regime, the numerical solution consists either in a quasi-2D condensate whose root-mean-square velocity adapts to reach $Ro \simeq 1$, or in a bursting behavior with long phases of quasi-2D motion and short and intermittent bursts of fully 3D turbulence. In any case, 2D motions play a key role in the dynamics.

The second situation is the one considered in the present study: when the forcing inputs energy directly into two-dimensional motion, then one can reach the regime of high-Reynolds number $Re$ and low-Rossby number $Ro$. However, for fixed system size and any given Reynolds number, we have proven rigorously that there is a critical value of the Rossby number under which the flow ends up in a purely 2D attractor, with no inertial waves at all.

The author thanks F. Moisy, W.R. Young, A. Alexakis, C.R. Doering, P.-P. Cortet and A. Campagne for insightful discussions. This research is supported by Labex PALM ANR-10-LABX-0039.

\appendix

\section{Bounds for body-forced 2D turbulent flows\label{AppendixA}}
In this appendix only, the $L_2$ norms refer to integration over the 2D periodic domain $(x,y) \in {\cal D}_2=[0, L]^2$. Like in the main body of the paper, we restrict attention to $Re \geq 2$.

\subsection{The analysis of Alexakis and Doering}
Following Alexakis and Doering, take the dot product of 2D Navier-Stokes equation (\ref{NS2D}) with ${ \bphi}$, integrate over the 2D domain,  divide by $L^2$ and time-average to get
\begin{eqnarray}
\nonumber F  & = & -\frac{\nu}{L^2} \la \int_{{\cal D}_2}  {\bf V} \cdot \Delta \bphi  \mathrm{d}^2 \textbf{x} \ra + \frac{1}{L^2} \la \int_{{\cal D}_2}  (({\bf V} \cdot \bnabla) {\bf V}) \cdot \bphi  \mathrm{d}^2 \textbf{x} \ra \\
 & \leq & \frac{\nu}{L^2} \| \Delta \bphi \|_2 \la \|  {\bf V} \|_2 \ra  +  \frac{1}{L^2} \| \bnabla \bphi \|_\infty \la \| {\bf V}\|_2^2 \ra  \lesssim  \frac{\nu U}{\ell^2}   +  \frac{U^2}{\ell}  \lesssim   \frac{U^2}{\ell} \label{boundF}
\end{eqnarray}
where we performed several integrations by parts before using the Cauchy-Schwarz, H\"older's and Jensen's inequalities, and we keep restricting attention to $Re \geq 2$. (\ref{boundF}) can be easily recast as an inequality between the Grashof and the Reynolds number:  
\begin{eqnarray}
Gr \lesssim Re^2 \, .\label{boundGr}
\end{eqnarray}

The bound on the enstrophy dissipation rate is obtained by taking the curl of (\ref{NS2D}), multiplying it by $\omega$, integrating over the domain and time-averaging. This leads to the enstrophy budget, which we divide by $\nu$ to obtain
\begin{eqnarray}
\nonumber \la \| \bnabla \omega \|_2^2 \ra & = & \frac{1}{\nu} \la \int_{{\cal D}_2} (\bnabla \times {\bf f})\cdot {\bf e}_z  \omega \mathrm{d}^2 \textbf{x} \ra \\
 & \leq & \frac{\| {\bf f} \|_2}{\nu} \la \| \bnabla \omega \|_2 \ra \leq \frac{\| {\bf f} \|_2}{\nu} \sqrt{\la \| \bnabla \omega \|_2^2 \ra} \lesssim \frac{F L}{\nu} \sqrt{\la \| \bnabla \omega \|_2^2 \ra} \, , 
\end{eqnarray}
where we integrated by parts before using the Cauchy-Schwarz and Jensen's inequalities. Squaring the last inequality and dividing by $\la \| \bnabla \omega \|_2^2 \ra$ leads to
\begin{eqnarray}
\la \| \bnabla \omega \|_2^2 \ra \lesssim \frac{F^2 L^2}{\nu^2} \, .\label{gradomegaAppF}
\end{eqnarray}
Similarly, the bound on the time-averaged enstrophy is obtained by considering the time-averaged energy budget divided by $\nu$
\begin{eqnarray}
 \la \| \omega \|_2^2 \ra & = & \frac{1}{\nu} \la \int_{{\cal D}_2} {\bf f} \cdot {\bf u} \mathrm{d}^2 \textbf{x} \ra 
\end{eqnarray}
Since ${\bf f}$ is divergence-free, we can write ${\bf f} = \bnabla \times \bxi$ where the vector potential $\bxi$ satisfies $\bnabla \cdot  \bxi = 0$, and because ${\bf f}$ has spatial period $\ell$, Poincar\'e's inequality yields $\| \bxi \|_2 \leq \ell \|  {\bf f} \|_2/(2 \pi)$. After an integration by parts,
\begin{eqnarray}
\la \| \omega \|_2^2 \ra & \leq & \frac{\| \bxi \|_2}{\nu}  \la \|  \omega \|_2 \ra \leq \frac{\| \bxi \|_2}{\nu}  \sqrt{\la \|  \omega \|_2^2 \ra}   \lesssim  \frac{F \ell L}{\nu} \sqrt{\la \|  \omega \|_2^2 \ra} \, , \\
\end{eqnarray}
and after squaring and dividing by $\la \| \omega \|_2^2 \ra$,
\begin{eqnarray}
\la \| \omega \|_2^2 \ra & \lesssim &  \frac{F^2 \ell^2 L^2}{\nu^2} \, .\label{omegaAppF}
\end{eqnarray}
The inequalities (\ref{boundEnstrophyF}) and (\ref{boundEDF}) correspond to (\ref{omegaAppF}) and (\ref{gradomegaAppF}), with an extra prefactor $H$, because $\| \dots  \|_2$ denotes the 2D $L_2$ norm throughout this appendix, and the 3D one in the main body of the paper.
Inserting the Poincar\'e inequality in (\ref{omegaAppF}) gives
\begin{eqnarray}
L^2 U^2 = \la \| {\bf u} \|_2^2 \ra \lesssim L^2 \la \| \omega \|_2^2 \ra & \lesssim &  \frac{F^2 \ell^2 L^4}{\nu^2} \, ,
\end{eqnarray}
hence
\begin{eqnarray}
Re \lesssim \frac{L}{\ell} Gr \, .
\end{eqnarray}
With $Re \geq 2$ and $Gr \geq 2$, this inequality and (\ref{boundGr}) lead to both
\begin{eqnarray}
\ln \left( Re \frac{L}{\ell} \right) \lesssim \ln \left( Gr \frac{L}{\ell} \right) \, \mbox{ and } \ln \left( Gr \frac{L}{\ell} \right) \lesssim \ln \left( Re \frac{L}{\ell} \right) \, . \label{boundln}
\end{eqnarray}

We now reproduce the analysis of Alexakis and Doering to obtain bounds in terms of the r.m.s. velocity $U$. Once again, the bound on the enstrophy dissipation rate is obtained by considering the enstrophy budget, which we divide by $\nu$ to obtain
\begin{eqnarray}
\nonumber \la \| \bnabla \omega \|_2^2 \ra & = & \frac{1}{\nu} \la \int_{{\cal D}_2} (\bnabla \times {\bf f})\cdot {\bf e}_z  \omega \mathrm{d}^2 \textbf{x} \ra \\
 & = & - \frac{1}{\nu} \la \int_{{\cal D}_2} {\bf V} \cdot \Delta {\bf f} \mathrm{d}^2 \textbf{x} \ra   \lesssim   \frac{1}{\nu} \| \Delta {\bf f} \|_2  \la \| {\bf V} \|_2 \ra  \lesssim \frac{F L^2 U}{\nu \ell^2} \lesssim \frac{L^2 U^3}{\ell^3 \nu} \, , \label{gradomegaApp}
\end{eqnarray}
where we made use of (\ref{boundF}) to get the last inequality.
The bound on the time-averaged enstrophy is obtained from an integration by parts together with the Cauchy-Schwarz inequality,
\begin{eqnarray}
 \la \| \omega \|_2^2 \ra & \leq & \la  \| {\bf V}  \|_2 \| \bnabla \omega  \|_2 \ra \leq L U \sqrt{\la \| \bnabla \omega \|_2^2 \ra} \lesssim \frac{L^2 U^2}{\ell^2} \sqrt{Re} \, . \label{omegaApp}
\end{eqnarray}
The inequalities (\ref{boundEnstrophy}) and (\ref{boundED}) correspond to (\ref{omegaApp}) and (\ref{gradomegaApp}), with an extra prefactor $H$, because $\| \dots  \|_2$ denotes the 2D $L_2$ norm throughout this appendix, and the 3D one in the main body of the paper.

In the case of  single-mode forcing, we have $\Delta {\bf f}(x,y) = - \frac{{\tilde c}_1}{\ell^2} {\bf f}(x,y)$, where ${\tilde c}_1$ is a dimensionless constant. The energy budget is obtained by dotting ${\bf V}$ into (\ref{NS2D}), integrating over the whole domain and time-averaging. This leads to
\begin{eqnarray}
 \la \| \omega \|_2^2 \ra =\frac{1}{\nu} \la \int_{{\cal D}_2} {\bf f} \cdot {\bf V} \mathrm{d}^2 \textbf{x} \ra = \frac{-\ell^2}{{\tilde c}_1 \nu} \la \int_{{\cal D}_2} \Delta {\bf f} \cdot {\bf V} \mathrm{d}^2 \textbf{x} \ra = \frac{\ell^2}{{\tilde c}_1} \la \| \bnabla \omega \|_2^2 \ra \, ,\label{tempKolm}
\end{eqnarray}
where the last equality originates from (\ref{gradomegaApp}).
Using the Cauchy-Schwarz inequality together with (\ref{tempKolm}),
\begin{eqnarray}
 \la \| \omega \|_2^2 \ra & \leq & \sqrt{\la  \| {\bf V}  \|_2^2 \ra}  \sqrt{ \la \| \bnabla \omega  \|_2^2 \ra} \lesssim  \frac{L U}{\ell}  \sqrt{ \la \|  \omega  \|_2^2 \ra } \,
\end{eqnarray}
hence
\begin{eqnarray}
 \la \| \omega \|_2^2 \ra   \lesssim  \frac{L^2 U^2}{\ell^2}  \mbox{ , and } \la \| \bnabla \omega \|_2^2 \ra=  \frac{{\tilde c}_1}{\ell^2} \la \| \omega \|_2^2 \ra \lesssim \frac{L^2 U^2}{\ell^4} \, .
\end{eqnarray}
These inequalities correspond to (\ref{boundEnstrophyKolmogorov}) and (\ref{boundEDKolmogorov}), up to a prefactor $H$ to go from the 2D $L_2$ norm of this appendix to the 3D one used in the main body of the paper.

\subsection{Bounds on the suprema in time of energy and enstrophy}

The base-flow ${\bf V}(x,y,t)$ lies on the attractor of the body-forced 2D Navier-Stokes equation. 
Consider a time $t_0>0$ for which the supremum of $\| {\bf V}\|_2^2$ is achieved (the proof could easily be adapted if the supremum were not achieved). The energy evolution equation at $t=t_0$ reads:
\begin{eqnarray}
\mathrm{d}_t \left. \left( \frac{\| {\bf V}\|_2^2}{2}\right) \right|_{t=t_0} = 0 =  \int_{{\cal D}_2} {\bf f} \cdot {\bf V} \mathrm{d}^2 \textbf{x}-\nu \left. \| \omega \|_2^2 \right|_{t=t_0}
\end{eqnarray}
Since ${\bf f}$ is divergence-free, we write ${\bf f} = \bnabla \times \bxi$ where the vector potential $\bxi$ satisfies $\bnabla \cdot  \bxi = 0$, and because ${\bf f}$ has spatial period $\ell$, Poincar\'e's inequality yields $\| \bxi \|_2 \leq \ell \|  {\bf f} \|_2/(2 \pi)$. After an integration by parts,
\begin{eqnarray}
\nu \left. \| \omega \|_2^2 \right|_{t=t_0} =  \int_{{\cal D}_2} {\bf f} \cdot {\bf V} \mathrm{d}^2 \textbf{x} \leq  \int_{{\cal D}_2} | \bxi |  |\omega| \mathrm{d}^2 \textbf{x} \leq \| \bxi \|_2 \left. \| \omega \|_2  \right|_{t=t_0} \lesssim L \ell F  \left. \| \omega \|_2  \right|_{t=t_0} \, ,
\end{eqnarray} 
hence
\begin{eqnarray}
\left. \| \omega \|_2 \right|_{t=t_0} \lesssim \frac{L \ell F}{\nu} \, .
\end{eqnarray} 
From Poincar\'e's inequality, $\| {\bf V} \|_2 \lesssim L \| \omega \|_2$, hence
\begin{eqnarray}
 \sup_t \| {\bf V} \|_2 =  \| {\bf V} \|_2|_{t=t_0}  \lesssim  \frac{L^2 \ell F}{\nu} \lesssim  \frac{L^2 U^2}{\nu} \, , \label{boundSupV}
\end{eqnarray}
where we used (\ref{boundF}) to get the last inequality. 

A very similar analysis performed on the enstrophy conservation equation leads to the following upper bound,
\begin{eqnarray}
\sup_t \| \omega \|_2 \lesssim \frac{F L^2}{ \nu}  \lesssim \frac{U^2 L^2}{\ell \nu}  \, . \label{boundSupomega}
\end{eqnarray}

\subsection{A (loose) bound on $\la \| \Delta \omega \|_2^2 \ra$}
Let us write the palinstrophy evolution equation,
\begin{eqnarray}
\mathrm{d}_t \left( \frac{\| \bnabla \omega \|_2^2}{2} \right) = \int_{{\cal D}_2}  ({\bf V}\cdot \bnabla \omega) \Delta \omega\mathrm{d}^2 \textbf{x} - \nu \| \Delta \omega \|_2^2 -  \int_{{\cal D}_2} (\bnabla \times \omega {\bf e}_z) \cdot \Delta {\bf f} \mathrm{d}^2 \textbf{x} \, ,
\end{eqnarray}
time-average it, and use successively H\"older's, the Cauchy-Schwarz and Young's inequalities,
\begin{eqnarray}
\nu \la \| \Delta \omega \|_2^2 \ra & \leq & \frac{\nu}{2} \la \| \Delta \omega \|_2^2 \ra +  \frac{1}{2\nu} \la \int_{{\cal D}_2}  |{\bf V}|^2  |\bnabla \omega |^2 \mathrm{d}^2 \textbf{x} \ra + \la \| {\bf V}\|_2 \ra \|  \Delta^2 {\bf f}\|_2 \, .\label{ineqA1}
\end{eqnarray}
To bound the quartic term, use the Cauchy-Schwarz and Ladyzhenskaya's inequalities \citep{Ladyzhenskaya},
\begin{eqnarray}
\la \int_{{\cal D}_2}  |{\bf V}|^2  |\bnabla \omega |^2 \mathrm{d}^2 \textbf{x} \ra & \leq & \la \left(\int_{{\cal D}_2}  |{\bf V}|^4 \mathrm{d}^2 \textbf{x} \int_{{\cal D}_2}  |\bnabla \omega |^4 \mathrm{d}^2 \textbf{x} \right)^{1/2} \ra \\
\nonumber & \lesssim &  \la \| {\bf V} \|_2 \| \omega \|_2 \| \bnabla \omega \|_2 \| \Delta \omega \|_2 \ra \\
\nonumber & \lesssim &  \sup_t \left(\| \omega \|_2\right)   \sup_t \left( \| {\bf V} \|_2 \right) \sqrt{\la \| \bnabla \omega \|_2^2 \ra } \sqrt{ \la \| \Delta \omega \|_2^2 \ra}  \\
\nonumber & \leq &  \frac{\nu^2}{2} \la \| \Delta \omega \|_2^2 \ra + \frac{c}{\nu^2}  \sup_t \left(\| \omega \|_2^2\right)   \sup_t \left( \| {\bf V} \|_2^2 \right) \la \| \bnabla \omega \|_2^2 \ra \, .
\end{eqnarray}
Inserting this inequality in (\ref{ineqA1}),
\begin{eqnarray}
\la \| \Delta \omega \|_2^2 \ra \lesssim \frac{FL^2 U}{\nu \ell^4}  +\frac{\sup_t \left(\| {\bf V} \|_2^2\right) \sup_t \left(\| \omega \|_2^2\right)}{ \nu^4} \la \| \bnabla \omega \|_2^2 \ra \, ,
\end{eqnarray}
and using the bounds (\ref{boundF}), (\ref{boundSupV}) and (\ref{boundSupomega}),
\begin{eqnarray}
\frac{\la \| \Delta \omega \|_2^2 \ra}{\la \| \bnabla \omega \|_2^2 \ra} \lesssim \left( \frac{ U^3 L^2}{\nu \ell^{5}\la \| \bnabla \omega \|_2^2 \ra}  +\frac{ U^8 L^{8}}{\ell^2 \nu^8} \right)  \, .
\end{eqnarray}
After another use of Poincar\'e's inequality, $\la \| \bnabla \omega \|_2^2 \ra \geq 16 \pi^4 U^2 / L^2$, so that
\begin{eqnarray}
\frac{L^2 \la \| \Delta \omega \|_2^2 \ra}{\la \| \bnabla \omega \|_2^2 \ra} \leq c \left( \frac{L}{\ell} \right)^{10} Re^8  \, .\label{boundarglog}
\end{eqnarray}
In the next subsection we are interested in the logarithm of this quantity which can be bounded as
 \begin{eqnarray}
\nonumber \ln \left( \frac{L^2 \la \| \Delta \omega \|_2^2 \ra}{\la \| \bnabla \omega \|_2^2 \ra} \right) &  \lesssim  & \ln \left(c \frac{L^{10}}{\ell^{10}} Re^8 \right) = \ln( c ) + 10 \ln \left( \frac{L}{\ell} \right) + 8 \ln Re \\
& \lesssim & \ln \left( Re \frac{L}{\ell} \right) \lesssim  \ln \left( Gr \frac{L}{\ell} \right) \, ,\label{boundarglog}
\end{eqnarray}
where we use $Re \geq 2$, and therefore $\ln Re \geq \ln 2 >0$, to remove $\ln (c)$. We have also used (\ref{boundln}) for the last step, assuming $Gr \geq 2$.

\subsection{A bound on $\la \| \bnabla {\bf V} \|_\infty^2 \ra$}
Let us consider an arbitrary cut-off wavenumber $Q>0$ and write
\begin{eqnarray}
 \| \bnabla {\bf V} \|_\infty & \leq & \sum\limits_{{\bf k}} k |{\bf V}_{\bf k}| = \sum\limits_{k < Q} k |{\bf V}_{\bf k}| + \sum\limits_{k \geq Q} k |{\bf V}_{\bf k}| \\
 \nonumber & \leq & \sqrt{\sum\limits_{k < Q} k^4 |{\bf V}_{\bf k}|^2} \sqrt{\sum\limits_{k < Q} \frac{1}{k^2}} + \sqrt{\sum\limits_{k \geq Q} k^6 |{\bf V}_{\bf k}|^2} \sqrt{\sum\limits_{k \geq Q} \frac{1}{k^4}}
\end{eqnarray}
Now sums that are independent of ${\bf V}$ can be bounded by integrals, with the following result
\begin{eqnarray}
\sum\limits_{k < Q} \frac{1}{k^2} & \lesssim & L^2 \ln (QL) \, , \\
\nonumber  \sum\limits_{k \geq Q} \frac{1}{k^4} & \lesssim & \frac{L^2}{Q^2} \, ,
\end{eqnarray}
hence
\begin{eqnarray}
 \| \bnabla {\bf V} \|_\infty \leq  \sum\limits_{{\bf k}} k |{\bf V}_{\bf k}| & \lesssim & \frac{1}{Q} \| \Delta \omega \|_2 + \sqrt{\ln (QL)} \| \bnabla \omega \|_2 \, ,\label{separateQ}
\end{eqnarray}
and after squaring, time-averaging and using Young's inequality,
\begin{eqnarray}
\la \| \bnabla {\bf V} \|_\infty^2 \ra \leq \la \left(  \sum\limits_{{\bf k}} k |{\bf V}_{\bf k}| \right)^2 \ra & \lesssim & \frac{1}{Q^2} \la \| \Delta \omega \|_2^2 \ra + \ln (QL) \la \| \bnabla \omega \|_2^2 \ra \, .
\end{eqnarray}
Upon picking $Q= \sqrt{\la \| \Delta \omega \|_2^2 \ra / \la \| \bnabla \omega \|_2^2 \ra}$ we obtain
\begin{eqnarray}
\la \| \bnabla {\bf V} \|_\infty^2 \ra \leq \la \left(  \sum\limits_{{\bf k}} k |{\bf V}_{\bf k}| \right)^2 \ra \lesssim \la \| \bnabla \omega \|_2^2 \ra \left[ 1+\ln \left( \frac{L^2 \la \| \Delta \omega \|_2^2 \ra}{ \la \| \bnabla \omega \|_2^2 \ra } \right) \right] \, .\label{boundtempinfty}
\end{eqnarray}
The combination of (\ref{boundtempinfty}) with the bound (\ref{boundarglog}) implies that there is a constant $\tilde{c}_2$ (depending on the shape of the forcing only), such that, for $Gr \geq 2$, 
\begin{eqnarray}
\la \| \bnabla {\bf V} \|_\infty^2 \ra \leq \la \left(  \sum\limits_{{\bf k}} k |{\bf V}_{\bf k}| \right)^2 \ra & \leq & \tilde{c}_2  \la \| \bnabla \omega \|_2^2 \ra \ln  \left( Gr  \frac{L}{\ell}  \right) \, .\label{boundGradVinf2}
\end{eqnarray}
We use this inequality to bound the cutoff wavenumber ${\cal K}$ chosen in (\ref{valueK}), which leads to (\ref{boundK}).

The same method can be used to determine a bound on $\la \|  {\bf V} \|_\infty^2 \ra $ as well, with the following result,
\begin{eqnarray}
\la \|  {\bf V} \|_\infty^2 \ra  \leq \la \left(  \sum\limits_{{\bf k}} |{\bf V}_{\bf k}| \right)^2 \ra & \lesssim & \la \|  \omega \|_2^2 \ra \ln \left(  Re \frac{L}{\ell}  \right) \lesssim \la \|  \omega \|_2^2 \ra \ln \left(  Gr \frac{L}{\ell}  \right) \, .\label{boundVinf2}
\end{eqnarray}
Recall that throughout this appendix $\| \dots \|_2$ refers to the $L_2$ norm in 2D.

\subsection{A bound on $\left< \sum\limits_{ {{\bf k}, k\leq 2{\cal K} ; s_{\bf k}}}   |\partial_t B_{s_{\bf k}}|     \right>$ \label{bounddt}}

Using the helical decomposition (\ref{helicalV}) together with (\ref{horizontalV}) and $i {\bf k} \times {\bf h}_{s_{\bf k}} = s_{\bf k} k {\bf h}_{s_{\bf k}}$, we can write a Fourier component of the vorticity as
\begin{eqnarray}
\nonumber \omega_{\bf k} & = & i {\bf k} \times {\bf V}_{\bf k} \cdot {\bf e}_z=  k \left[B_+({\bf k}) {\bf h}_+({\bf k})-B_-({\bf k}) {\bf h}_-({\bf k}) \right] \cdot {\bf e}_z=  k B_+({\bf k}) [{\bf h}_+({\bf k})+{\bf h}_-({\bf k})] \cdot {\bf e}_z\, \\
 & = & - \sqrt{2}  k B_+({\bf k}) = \sqrt{2}  k B_-({\bf k}) \, ,
\end{eqnarray}
where we use the same convention as in (\ref{FourierV}) for all the Fourier series considered in this section. 
We therefore obtain
\begin{eqnarray}
 \sum\limits_{\substack{{{\bf k} ; k \leq 2 {\cal K}} \\ s_{\bf k} }} |\partial_t B_{s_{\bf k}} |   =  \sqrt{2} \sum\limits_{{\bf k} ; k \leq 2 {\cal K}} \frac{| \partial_t \omega_{\bf k}  |}{k} \, ,
\end{eqnarray}
the right hand-side of which can be bounded using the 2D vorticity evolution equation written for a single Fourier component,
\begin{eqnarray}
| \partial_t \omega_{\bf k}  | \leq |{\bf k}\cdot \{\omega {\bf V}\}_{\bf k}  | + \nu k^2 |\omega_{\bf k}| + |{\bf k}\times {\bf f}_{\bf k}| \, .
\end{eqnarray}
In this expression, $ \{\omega {\bf V}\}_{\bf k}$ denotes the Fourier amplitude of the product $\omega {\bf V}$.
After dividing by $k$, summing over ${\bf k}$ such that $k \leq 2 {\cal K}$, and time-averaging,
\begin{eqnarray}
\frac{1}{\sqrt{2}}  \la  \sum\limits_{\substack{{{\bf k} ; k \leq 2 {\cal K}} \\ s_{\bf k} }} |\partial_t B_{s_{\bf k}} | \ra  & \leq & \la \sum\limits_{{\bf k} ; k \leq 2 {\cal K}}  | \{\omega {\bf V}\}_{\bf k}  | \ra + \nu \la \sum\limits_{{\bf k} ; k \leq 2 {\cal K}} k |\omega_{\bf k}| \ra \label{tempdt}\\
\nonumber & & + \sum\limits_{{\bf k} ; k \leq 2 {\cal K}}  |{\bf f}_{\bf k}| \, .
\end{eqnarray}
The contribution from the nonlinear term in (\ref{tempdt}) is bounded according to
\begin{eqnarray}
\sum\limits_{{\bf k} ; k \leq 2 {\cal K}}  | \{\omega {\bf V}\}_{\bf k}  | & \leq & \sum\limits_{{\bf k} ; k \leq 2 {\cal K}}  \sum\limits_{{\bf p}}  |\omega_{\bf p}| |{\bf V}_{{\bf k}-{\bf p}} |   \leq   \left( \sum\limits_{{\bf p}}  |\omega_{\bf p}| \right) \left( \sum\limits_{{\bf k}}  |{\bf V}_{\bf k}| \right)\, ,
\end{eqnarray}
and after time-averaging and using the Cauchy-Schwarz inequality,
\begin{eqnarray}
\la \sum\limits_{{\bf k} ; k \leq 2 {\cal K}}  | \{\omega {\bf V}\}_{\bf k}  | \ra & \leq & \sqrt{ \la \left( \sum\limits_{{\bf p}} p |{\bf V}_{\bf p}| \right)^2 \ra} \sqrt{ \la \left( \sum\limits_{{\bf k}}  |{\bf V}_{\bf k}| \right)^2 \ra }\, .
\end{eqnarray}
and using the bounds (\ref{boundVinf2}) and (\ref{boundGradVinf2}),
\begin{eqnarray}
 \la \sum\limits_{{\bf k} ; k \leq 2 {\cal K}}  | \{\omega {\bf V}\}_{\bf k}  | \ra  & \lesssim & \sqrt{ \la \|  \omega \|_2^2 \ra \la \| \bnabla \omega \|_2^2 \ra} \ln \left(  Gr  \frac{L}{\ell}  \right) \, .\label{NLtermapp}
\end{eqnarray}
The viscous contribution is bounded according to
\begin{eqnarray}
\nonumber \nu \la \sum\limits_{{\bf k} ; k \leq 2 {\cal K}} k |\omega_{\bf k}| \ra & \lesssim & \nu {\cal K} \la  \sum\limits_{{\bf k} ; k \leq 2 {\cal K}}  |\omega_{\bf k}| \ra \lesssim \nu {\cal K} \sqrt{\ln ({\cal K}L)} \sqrt{\la \| \bnabla \omega \|_2^2 \ra}  \\
& \lesssim & H \la \| \bnabla \omega \|_2^2 \ra \ln \left(Gr \frac{L}{\ell} \right) \, , \label{viscoustermapp}
\end{eqnarray}
where we used the Cauchy-Schwarz inequality, and the bounds (\ref{boundK}) and (\ref{boundlnKL}) on ${\cal K}$ and $\ln ({\cal K}L)$, recalling that $\| \dots \|_2$ denotes the 2D $L_2$ norm in this appendix.

The forcing term is bounded as follows,
\begin{eqnarray}
 \sum\limits_{{\bf k} ; k \leq 2 {\cal K}}  |{\bf f}_{\bf k}| & \lesssim &  F \sum\limits_{{\bf k} ; k \leq 2 {\cal K}}  |{\bphi}_{\bf k}| \lesssim F \sqrt{\sum\limits_{{\bf k} } k^4 |{\bphi}_{\bf k}|^2}  \sqrt{\sum\limits_{{\bf k} } \frac{1}{k^4}} \lesssim FL \| \Delta \bphi \|_2 \lesssim \frac{F L^2}{\ell^2} \, . \quad \label{forcingtermapp} 
\end{eqnarray}

Using the bounds (\ref{NLtermapp}), (\ref{viscoustermapp}) and (\ref{forcingtermapp}) on the three terms on the right-hand-side of (\ref{tempdt}), together with the bound (\ref{boundK}) on the cut-off wavenumber ${\cal K}$, we obtain
\begin{eqnarray}
\nonumber  \la  \sum\limits_{\substack{{{\bf k} ; k \leq 2 {\cal K}} \\ s_{\bf k} }} |\partial_t B_{s_{\bf k}} |   \ra   \lesssim   
\frac{F L^2 }{\ell^2} +  \left( \sqrt{\la \|  \omega \|_2^2 \ra \la \| \bnabla \omega \|_2^2 \ra} + H \la \| \bnabla \omega \|_2^2 \ra \right) \ln \left(Gr \frac{L}{\ell} \right) \, . \\
\end{eqnarray}
Recall that throughout this appendix $\| \dots \|_2$ refers to the $L_2$ norm in 2D.

\section{Bounds for absolute two-dimensionalization \label{secBabs}}
 
 The term ${\cal T}_1^{(n)}$ is bounded using the Cauchy Schwarz inequality,
\begin{eqnarray}
|{\cal T}_1^{(n)}| & \lesssim & \frac{\Kabs^2}{ T } \int_{t=T_0}^{t=T_0+T} \left[ \sum\limits_{\substack{{\bf p}+{\bf q}+{\bf k}=0 \\ k_z=\frac{n \pi}{H};k \leq \Kabs; p \leq \Kabs\\ s_{\bf p}; {s_{\bf q}}; s_{\bf k} }}  |\partial_t(B_{s_{\bf q}})|   \frac{|b_{s_{\bf k}}| | b_{s_{\bf p}}|}{\| {\bf v} \|_2^2}    \right] \mathrm{d}t \\
\nonumber & \lesssim &  \frac{\Kabs^2}{ T } \int_{t=T_0}^{t=T_0+T}   \sum\limits_{\substack{{\bf q}; q\leq 2 \Kabs  \\ s_{\bf q}}}   |\partial_t(B_{s_{\bf q}})|        \left[   \frac{ \sum\limits_{\substack{{\bf k} , k_z=\frac{n \pi}{H}}}   |{\bf v}_{\bf k}|^2}{\| {\bf v} \|_2^2}    \right] \mathrm{d}t  \, .
\end{eqnarray}
and therefore
\begin{eqnarray}
\sum_{n \in \mathbb{Z}_{\neq 0}} |{\cal T}_1^{(n)}| & \lesssim & \frac{\Kabs^2}{L^2 H T}  \int_{t=T_0}^{t=T_0+T}   \sum\limits_{\substack{{\bf q}; q\leq 2 \Kabs  \\ s_{\bf q}}}   |\partial_t(B_{s_{\bf q}})|       \mathrm{d}t    \\
\nonumber & \lesssim & \frac{\Kabs^4}{H} \sup_{t>T_0; q\leq 2 \Kabs; s_{\bf q}}   |\partial_t(B_{s_{\bf q}})|
\end{eqnarray}

$\partial_t(B_{s_{\bf q}})$ is given by (\ref{eqBskabs}). For $q \leq 2 \Kabs$, the terms on the right-hand side of this equation can be easily bounded by $\Omega$-independent and time-independent quantities using (\ref{thBMN}), (\ref{thBMNinfty}) and the Cauchy-Schwarz inequality. With $\Kabs$ bounded in (\ref{boundKabs}), we finally get 
\begin{eqnarray}
\sum_{n \in \mathbb{Z}_{\neq 0}} |{\cal T}_1^{(n)}| & \leq & C \, ,\label{BT1absapp}
\end{eqnarray}
where $C$ still denotes the generic $\Omega$-independent and time-independent quantity (it may depend on $F$, $\nu$, $H$, $\ell$, $L$, and the shape of the forcing).

We now bound the term ${\cal T}_2^{(n)}$ using the Cauchy-Schwarz inequality
\begin{eqnarray}
|{\cal T}_2^{(n)}| & \lesssim & \frac{\Kabs^2}{ T } \int_{t=T_0}^{t=T_0+T} \left[ \sum\limits_{\substack{{\bf p}+{\bf q}+{\bf k}=0 \\ k_z=\frac{n \pi}{H};k \leq \Kabs; p \leq \Kabs\\ s_{\bf p}; {s_{\bf q}}; s_{\bf k} }}  |B_{s_{\bf q}}|   \frac{|b_{s_{\bf k}} \partial_t b_{s_{\bf p}} +b_{s_{\bf p}} \partial_t b_{s_{\bf k}}|}{\| {\bf v} \|_2^2}    \right] \mathrm{d}t \\
\nonumber & \lesssim &  \frac{\Kabs^2}{ T } \int_{t=T_0}^{t=T_0+T}   \sum\limits_{\substack{{\bf q}; q\leq 2 \Kabs  \\ s_{\bf q}}}   |B_{s_{\bf q}}|      \frac{  \sqrt{\sum\limits_{\substack{k \leq \Kabs , k_z=\frac{n \pi}{H} \\ s_{\bf k}}}    | b_{s_{\bf k}}|^2 }   \sqrt{\sum\limits_{\substack{k \leq \Kabs , k_z=\frac{n \pi}{H} \\ s_{\bf k}}}   |\partial_t b_{s_{\bf k}}|^2 } } {\| {\bf v} \|_2^2}    \mathrm{d}t  \\
\nonumber & \lesssim &  \frac{\Kabs^4 L^2}{ T }  \sup_{t>T_0; q\leq 2 \Kabs; s_{\bf q}}   ( |B_{s_{\bf q}}|   )   \int_{t=T_0}^{t=T_0+T}   \frac{  \sqrt{\sum\limits_{\substack{k \leq \Kabs , k_z=\frac{n \pi}{H} \\ s_{\bf k}}}    | b_{s_{\bf k}}|^2 }   \sqrt{\sum\limits_{\substack{k \leq \Kabs, k_z=\frac{n \pi}{H} \\ s_{\bf k}}}   |\partial_t b_{s_{\bf k}}|^2 } } {\| {\bf v} \|_2^2}    \mathrm{d}t  \, ,
\end{eqnarray}
and after summing over $n$ and using the Cauchy-Schwarz inequality
\begin{eqnarray}
\sum_{n \in \mathbb{Z}_{\neq 0}}  |{\cal T}_2^{(n)}|  & \lesssim &  \frac{\Kabs^4 L}{\sqrt{H} T }  \sup_{t>T_0; q\leq 2 \Kabs; s_{\bf q}}   ( |B_{s_{\bf q}}|   )   \int_{t=T_0}^{t=T_0+T}   \frac{    \sqrt{\sum\limits_{\substack{    k \leq \Kabs   \\ s_{\bf k}}}   |\partial_t b_{s_{\bf k}}|^2 } } {\| {\bf v} \|_2}    \mathrm{d}t  \, .
\end{eqnarray}
Using (\ref{thBMNinfty}) we obtain
\begin{eqnarray}
\sum_{n \in \mathbb{Z}_{\neq 0}}  |{\cal T}_2^{(n)}|  & \lesssim &  \frac{C}{T }  \int_{t=T_0}^{t=T_0+T}   \frac{    \sqrt{\sum\limits_{\substack{    k \leq \Kabs   \\ s_{\bf k}}}   |\partial_t b_{s_{\bf k}}|^2 } } {\| {\bf v} \|_2}    \mathrm{d}t  \, .\label{T2temp}
\end{eqnarray}
We now bound the sum inside the square-root using equation (\ref{eqbskabs}) and the Cauchy-Schwarz inequality:
\begin{eqnarray}
\sqrt{\sum\limits_{\substack{    k \leq \Kabs   \\ s_{\bf k}}}   |\partial_t b_{s_{\bf k}}|^2}  & \leq & \sum\limits_{\substack{    k \leq \Kabs   \\ s_{\bf k}}}   |\partial_t b_{s_{\bf k}}| \\
\nonumber & \leq & \sum\limits_{\substack{    k \leq \Kabs   \\ s_{\bf k}}}   \nu k^2 |  b_{s_{\bf k}} | + \sum\limits_{\substack{    k \leq \Kabs   \\ s_{\bf k}}}  \sum\limits_{{\bf q}; s_{\bf q}; s_{\bf p}}   | b_{s_{\bf p}} (-{\bf q}-{\bf k}) | (| b_{s_{\bf q}}| + | B_{s_{\bf q}} |) (q+  |{\bf q} +{\bf k}|) \\
\nonumber & \leq & \sqrt{\sum\limits_{\substack{    k \leq \Kabs   \\ s_{\bf k}}} \nu^2 k^4}  \sqrt{ \sum\limits_{\substack{    k \leq \Kabs   \\ s_{\bf k}}} |  b_{s_{\bf k}} |^2} \\
\nonumber & & +  \left(\sum\limits_{\substack{    k \leq \Kabs   \\ s_{\bf k}}} 1 \right) \sqrt{   \sum\limits_{{\bf q}; s_{\bf q}}   (2q+\Kabs)^2 (2 | b_{s_{\bf q}}|^2 + 2 | B_{s_{\bf q}} |^2)  } \sqrt{   \sum\limits_{{\bf q}; s_{\bf q}}   | b_{s_{\bf q}}|^2}  \, , 
\end{eqnarray}
which using (\ref{thBMNinfty}) and (\ref{boundKabs}) yields
\begin{eqnarray}
\sqrt{\sum\limits_{\substack{    k \leq \Kabs   \\ s_{\bf k}}}   |\partial_t b_{s_{\bf k}}|^2}  & \leq &C  \sqrt{   \sum\limits_{{\bf k}; s_{\bf k}}   | b_{s_{\bf k}}|^2} \\
\nonumber & \leq & C \| {\bf v} \|_2 \, .
\end{eqnarray}
Inserting this inequality in (\ref{T2temp}) finally gives
\begin{eqnarray}
\sum_{n \in \mathbb{Z}_{\neq 0}}  |{\cal T}_2^{(n)}|  & \leq &  C \, .\label{BT2absapp}
\end{eqnarray}
We now take care of ${\cal T}_3^{(n)}$ using (\ref{eqv2}), which holds for ${\bf v}$ of arbitrary amplitude, and the Cauchy-Schwarz inequality:
\begin{eqnarray}
\nonumber \sum_{n \in \mathbb{Z}_{\neq 0}} | {\cal T}_3^{(n)} | & \lesssim &  \frac{1}{T } \int_{t=T_0}^{t=T_0+T} \left[ \sum\limits_{\substack{{\bf p}+{\bf q}+{\bf k}=0 \\ k \leq {\Kabs}; p \leq {\Kabs}\\  s_{\bf p}; s_{\bf q}; s_{\bf k} }} | B_{s_{\bf q}}|   |b_{s_{\bf p}}|  |b_{s_{\bf k}}|    { k p} \, |{g}_{ {\bf q} {\bf k} {\bf p}}^{ s_{\bf q} s_{\bf k} s_{\bf p}}|  \frac{ \| \bnabla {\bf V} \|_\infty  \| {\bf v} \|_2^2   + \nu \|\bnabla \textbf{v}\|_2^2}{\| {\bf v} \|_2^4} \right] \mathrm{d}t \\
\nonumber & \lesssim &       \frac{\Kabs^2}{L^2 H T } \int_{t=T_0}^{t=T_0+T}    \sum\limits_{\substack{{\bf q}; q\leq 2 \Kabs  \\ s_{\bf q}}}   |B_{s_{\bf q}}| \left(   \| \bnabla {\bf V} \|_\infty  + \nu \frac{ \|\bnabla \textbf{v}\|_2^2}{ \| \textbf{v}\|_2^2}    \right) \mathrm{d}t \\
 & \lesssim &  C_3 + \frac{C_4}{T}  \int_{t=T_0}^{t=T_0+T}  \frac{ \|\bnabla \textbf{v}\|_2^2}{ \| \textbf{v}\|_2^2}    \mathrm{d}t \, ,\label{BT3absapp}
\end{eqnarray}
where we used (\ref{thBMN}) and (\ref{boundKabs}) for the last step.

${\cal T}_4^{(n)}$ is given by
\begin{eqnarray}
\sum_{n \in \mathbb{Z}_{\neq 0}} {\cal T}_4^{(n)} & = &\frac{1}{T} \left[ \sum\limits_{\substack{{\bf p}+{\bf q}+{\bf k}=0 \\ k \leq {\Kabs}; p \leq {\Kabs} ; \sigma_{s_{\bf k}}+\sigma_{s_{\bf p}} \neq 0\\ s_{\bf p}; s_{\bf q}; s_{\bf k} }} B_{s_{\bf q}} \frac{b_{s_{\bf k}} b_{s_{\bf p}}}{\| {\bf v} \|_2^2}  e^{i(\sigma_{s_{\bf k}}+\sigma_{s_{\bf p}})t}  (-i) s_{\bf k} s_{\bf p} k p \, {g}_{ {\bf q} {\bf k} {\bf p}}^{*   s_{\bf q} s_{\bf k} s_{\bf p}}  \right]_{t=T_0}^{t=T_0+T} .
\end{eqnarray}
Isolating the sum over ${\bf q}$ and using the Cauchy-Schwarz inequality for the sum over ${\bf k}$ and ${\bf p}$,
\begin{eqnarray}
\sum_{n \in \mathbb{Z}_{\neq 0}} | {\cal T}_4^{(n)} |& \lesssim &\frac{\Kabs^2}{L^2 H T} \sum\limits_{ q\leq 2 \Kabs; s_{\bf q}} \left[ | B_{s_{\bf q}}|(T_0) + | B_{s_{\bf q}}|(T_0+T) \right] \\
\nonumber & \leq & \frac{C}{T} \, ,
\end{eqnarray}
where the last step consists in using the Cauchy-Schwarz inequality for the sum over ${\bf q}$, followed by (\ref{thBMN}) and (\ref{boundKabs}). This shows that
\begin{eqnarray}
\lim_{T \to \infty} \sum_{n \in \mathbb{Z}_{\neq 0}} | {\cal T}_4^{(n)} |= 0 \, .\label{BT4absapp}
\end{eqnarray}
Expressions (\ref{BT1absapp}), (\ref{BT2absapp}), (\ref{BT3absapp}) and (\ref{BT4absapp}) imply (\ref{boundT1abs})-(\ref{boundT4abs}).

\end{document}